\documentclass{pas}

\usepackage{multirow}
\usepackage{subcaption}

\definecolor{mygreen}{RGB}{0,128,0}

\begin{document}

\lefttitle{Publications of the Astronomical Society of Australia}
\righttitle{Cambridge Author}

\jnlPage{1}{4}
\jnlDoiYr{2026}
\doival{10.1017/pasa.xxxx.xx}

\articletitt{Research Paper}

\title{KL Dra as a Benchmark Laboratory for Accretion-Disk Physics: Constraints from TESS and Ground-Based Surveys}



\author{\gn{Luis E.} \sn{Salazar Manzano}$^{1,2}$,
        \gn{Liliana E.} \sn{Rivera Sandoval}$^{2,3}$,
        \gn{Jean-Marie} \sn{Hameury}$^{4}$,
        \gn{Craig O.} \sn{Heinke}$^{5}$,
        \gn{Iwona} \sn{Kotko}$^{6}$,
        \gn{Thomas J.} \sn{Maccarone}$^{7}$, and
        \gn{Manuel} \sn{Pichardo Marcano}$^{8}$
        }

\affil{$^1$Department of Astronomy, University of Michigan, Ann Arbor, MI 48109, USA\\
       $^2$Department of Physics and Astronomy, University of Texas Rio Grande Valley, Brownsville, TX 78520, USA\\
       $^3$South Texas Space Science Institute, University of Texas Rio Grande Valley, Brownsville, TX 78520, USA\\
       $^4$Universit\'e de Strasbourg, CNRS, Observatoire astronomique de Strasbourg, UMR 7550, 67000 Strasbourg, France\\
       $^5$Physics Dept., University of Alberta, Edmonton, AB, T6G 2E1, Canada\\
       $^6$Astronomical Observatory, Jagiellonian University, ulica Orla 171, 30-244 Krak\'ow, Poland\\
       $^7$Department of Physics \& Astronomy, Texas Tech University, Box 41051, Lubbock, TX 79409-1051, USA\\
       $^8$Universidad Nacional Aut\'onoma de M\'exico, Instituto de Astronom\'ia, Ciudad Universitaria, 04510 Ciudad de M\'exico, Mexico
       }


\corresp{Luis E. Salazar Manzano, Email: lesamz@umich.edu}

\citeauth{}

\history{(Received xx xx xxxx; revised xx xx xxxx; accepted xx xx xxxx)}

\begin{abstract}
We present the longest-term optical analysis of the AM CVn system KL Dra using $\sim11$ years of monitoring from TESS and wide-field ground-based surveys. The continuous TESS coverage allows us to characterise its frequent outbursts with unprecedented detail, providing the first comprehensive study of an AM CVn during outbursts and enabling detailed modelling of these systems. The superoutbursts in KL Dra generally include a precursor, and are followed by a series of rebrightenings after which a sequence of 3-4 large amplitude normal outbursts is observed. We fit parametric profiles to each superoutburst component (precursor, rise to plateau, plateau, decay), to rebrightenings, and to normal outbursts, which let us quantify every high state feature and investigate correlations with the system's long term supercyle evolution. Our continuous coverage reveals an average value for the supercycles, superoutbursts and  normal outbursts of $60.4 \pm 0.1$ d, $5.67\pm0.03$ d and $1.17 \pm0.01$ d, respectively. The supercycle duration may be correlated with the rebrightenings duration and superoutburst amplitude, and anticorrelated with the plateau length. Within a supercycle, normal outbursts grow in amplitude and duration, and the first normal outburst is usually highly asymmetric, while subsequent normal outbursts are more symmetric. We detected superhumps in TESS superoutbursts but not in the rebrightenings or normal outbursts. We interpret the results within the disc instability model, considering additional effects, such as changes in the donor mass transfer rate. 
\end{abstract}

\begin{keywords}
AM CVn stars, Accretion, Time domain astronomy
\end{keywords}

\maketitle

\section{Introduction} \label{sec:intro}

AM CVn systems\footnote{Also known as ultra-compact white dwarf binaries, interacting binary white dwarfs, or helium cataclysmic variables.} are a relatively rare class of accreting compact objects. They consist of a white dwarf accreting He-rich material from a degenerate or semi-degenerate donor star. These systems exhibit distinctive properties, such as the absence of hydrogen in their spectra, and their short orbital periods ranging from $\sim$5 to $\sim$70 min \citep{Solheim:2010,Green:2025A&A...700A.107G}. The evolutionary pathways leading to the formation of AM CVns must involve one or two phases of unstable mass transfer that culminate in a common envelope event. This fine tuning of their evolution makes the study of AM CVn systems important for binary stellar evolution theory. They are also potential supernovae progenitors \citep{2007Bildsten}, and anticipated sources for future low-frequency space-based gravitational wave observatories such as LISA \citep{Amaro-Seoane2023}. In addition, AM~CVn systems provide excellent opportunities to investigate accretion physics.

AM CVns are commonly classified as persistent or transient (outbursting) systems. Shorter orbital period AM CVns have more massive white dwarf donors (since the donors are degenerate), and have higher mass transfer rates than longer-period systems \citep{2004Marsh}. The shorter-period systems ($P_\mathrm{orb} \lesssim $ 20 min) have higher mass transfer rates and physically smaller discs than the longer-period systems, meaning that the short-period systems are able to keep their accretion discs fully ionised continuously, and thus stay persistently bright. Longer-period systems show an ionisation instability that leads to transient behaviour \citep{2008Lasota}. In these transient systems (20 $ \lesssim  P_\mathrm{orb} \lesssim $ 60 min) the accretion disc is unstable, oscillating between high and low states. They exhibit normal outbursts (NOs) and superoutbursts (SOs), and their spectra display different features in high and low states \citep{2013Levitan}. 

The physics behind these transient episodes in AM~CVns is often thought to be analogous to those of SU UMa class of dwarf novae \citep[see][and references therein]{Hameury:2020}, and as such has been modelled using a modified version of the Disk Instability Model (DIM) that includes enhanced mass transfer. The DIM proposes that the transition from quiescence to outburst is triggered by a thermal instability once the ionisation temperature is reached, enhancing the viscosity which transports the accumulated matter in the disc onto the white dwarf \citep{Hameury:2020}. The disc becomes thermally unstable when local conditions place an annulus in the range bounded by the critical surface densities $\Sigma_{\mathrm{min}}(r)$ and $\Sigma_{\mathrm{max}}(r)$. On the DIM S-curves, these thresholds are equivalently expressed as critical temperatures or critical accretion rates. However the traditional DIM has troubles explaining SOs.  To account for the longer and brighter phenomenon, the tidal-thermal instability (TTI) model extends the DIM by introducing a tidal instability that develops when the outer edge of the accretion disc reaches the 3:1 resonance radius \citep{2005Osaki}. This produces a sudden increase in tidal torque, forming an eccentric, precessing disc that enhances angular momentum transport. Yet recent observations \citep{2020RS,RiveraSandoval:2021, RiveraSandoval:2022} indicate that the DIM or TTI mechanism alone cannot fully explain the transient behaviour in AM~CVns. The enhanced mass transfer (EMT) model offers a complementary scenario in which variations in the mass transfer rate from the secondary star drive the binary's transient behaviour and when combined with the DIM can help explain the origin of SOs \citep{Kotko:2012A&A}. Although the origin of EMT is still uncertain, a possible explanation includes irradiation of the donor star by a tilted/warped disc. At the end of many SOs, AM CVn systems exhibit rebrightenings, a series of low-amplitude short-duration outbursts. In dwarf novae they have been described to be the result of the reflection of a cooling front, mass transfer variations, irradiation of the secondary, disc truncation or even residual viscosity within the disc \citep{Osaki:1997PASJ, Kotko:2012A&A, Kato:2015PASJ, 2021Hameury}.

Advancing our understanding of accretion models and assessing the relevance of additional ingredients critically depend on the availability of high quality data capable of constraining the large number of variables involved. In this context, KL Dra represents an exceptional laboratory for accretion studies for several reasons. Its declination of $\sim60^\circ$ makes it circumpolar for many northern observatories, ensuring ground-based coverage during most of the year. Its high ecliptic latitude also positions it near the ecliptic pole, within the overlap of several Transiting Exoplanet Survey Satellite (TESS) pointings. The synergy between its excellent observability and its frequent outburst activity results in one of the prime targets for investigating accretion processes.

KL Dra is an AM~CVn system with a photometric 25 min orbital period \citep{2002Wood}. It was first identified as the supernova candidate SN 1998di, due to its apparent proximity to the nucleus of a galaxy and its absence in previous surveys and catalogues \citep{Schwartz:1998}. Within the  month of discovery, spectroscopic observations identified He absorption lines, which led to its identification as an AM CVn system given the similarity to the spectrum of CR Boo \citep{Jha:1998}. Since then KL Dra has been subject to extensive theoretical and observational work.

The general optical properties that reflect KL Dra's outburst behaviour have been constrained by a combination of long term monitoring from the ground and high cadence observations from space. Early ground-based studies identified a recurrence time of $\sim60$ days and reported SO durations of $\sim15$ days with $\sim3$ mag amplitudes  \citep{Ramsay:2010, Ramsay:2012_2}. More recent observations by TESS have refined our understanding of KL~Dra's behaviour, confirming the existence of NOs \citep{Duffy:2021} (previously hinted at from the ground by \citealt{Ramsay:2010, Ramsay:2012_2}), revealing the presence of precursors \citep{Duffy:2021}, and clearly separating KL Dra's SOs from rebrightenings \citep{Pichardo:2021}. 

To have a comprehensive picture of the optical behaviour of KL~Dra and understand the different physical processes at play through accretion disc models, we require high cadence data and long term monitoring to probe the different variability regimes. In this work we exploit the continuous and high cadence observations of TESS, together with the long-term coverage of TESS and ground-based surveys in a synergistic manner to construct the most detailed study of the transient behaviour of KL Dra. In Section~\ref{sec:obser} we introduce the datasets used in this work, and in Section~\ref{sec:analy} our light curve analysis and fits to the outbursts. In Section~\ref{sec:res} we present our results for the different stages of the supercycles (SCs), and in Section~\ref{sec:dis} we discuss them in the context of the different models for binary accretion through accretion discs. We finish the manuscript with a summary and our conclusions in Section~\ref{sec:concl}.

\section{Observations} \label{sec:obser}

We used high-cadence observations provided by TESS. We complement TESS data with the long-term coverage provided by ground-based surveys such as the Zwicky Transient Facility (ZTF), the All-Sky Automated Survey for Supernovae (ASAS-SN) and the Asteroid Terrestrial-impact Last Alert System (ATLAS). By combining these datasets we are able to measure and characterise the variability on the scale of days for the SOs, and on the scale of months / years for SCs. When the original timestamps were not in UTC, we converted them to the Universal Coordinated Time (UTC) scale, and all timestamps reported throughout this work are given as Julian Dates (JD).

\subsection{TESS} \label{subsec:space}

TESS is a NASA's space-based observatory launched in 2018. It aims to find Earth-sized exoplanets around nearby bright stars \citep{Ricker:2014}. In its second extension, it has covered $95\%$ of the sky using a ``Stare and Step" strategy, observing 97 sectors for 27.4 days. TESS produces Full Frame Images (FFIs) and higher-cadence Target Pixel Files (TPFs) with a pixel scale of 21~$''/$pixel.

The primary data sets for this work are the 20 sec TPF data requested within the framework of our TESS General Investigator (GI) programs for northern targets (PI: Rivera Sandoval). KL Dra is located in the TESS continuous viewing zone because of its high ecliptic latitude, where the overlap among sectors results in continuous coverage of up to one year in each cycle. The exceptions are cycle 5 and 7, due to the alternating TESS observing strategy. We used FFIs with cadence of 200 sec in sectors 80, 81 and 84, and TPF with cadence of 120 sec from sector 85. For completeness, we have also used data from cycle 2, already reported in \cite{Duffy:2021} and \cite{Pichardo:2021}, which, despite its lower cadence (30 min FFIs), are included in this study to quantitatively characterise the outbursting features and the long-term behaviour of the binary. A total of 41 sectors are analysed in this work, see Table \ref{tab:tessdata} for details regarding the TESS data used. TESS images were downloaded and manipulated using the \texttt{Lightkurve} package \citep{Barentsen:2021}.

\begin{table}[]
\centering
\caption{Details for the TESS data used in this work.}
\label{tab:tessdata}

\begin{tabular}{cccc}
\hline
\hline
\textbf{Cycle} & \textbf{Sectors} & \textbf{Cadence} & \textbf{Images} \\
\hline
2 & 14, 16 - 26 & 30 min & FFI \\
4 & 40 - 41, 47 - 55 & 20 sec & TPF \\
5 & 56 - 60 & 20 sec & TPF \\
6 & 73 - 79, 82, 83 & 20 sec & TPF \\
6 & 80, 81 & 200 sec & FFI \\
7 & 84 & 200 sec & FFI \\
7 & 85 & 120 sec & TPF \\
\hline
\end{tabular}
\end{table}

\subsection{ZTF, ASAS-SN, and ATLAS} \label{subsec:ground}

ZTF \citep{Graham:2019, Bellm:2019}, ASAS-SN \citep{Shappee:2014, Kochanek:2017} and ATLAS \citep{Tonry:2018, Smith:2020PASP, Heinze:2018AJ} are optical time-domain surveys that monitor the sky for transient and variable phenomena. We use all publicly available data for KL Dra from these surveys. Specifically, for ZTF, we use the most recent DR23 data release \citep{Masci:2019PASP}. For ASAS-SN, we obtained the light curve data using the Sky Patrol service with the aperture photometry method \citep{Hart:2023arXiv}, while for ATLAS, we used their forced photometry server \citep{Shingles:2021TNSAN}. The ASAS-SN and ATLAS data were retrieved as of January 2025, while the ZTF data release used has a November 2024 cutoff.

\section{Data Analysis} \label{sec:analy}

\begin{figure*}
    \centering
    \begin{minipage}{\textwidth}
        \centering
        \includegraphics[width=\linewidth]{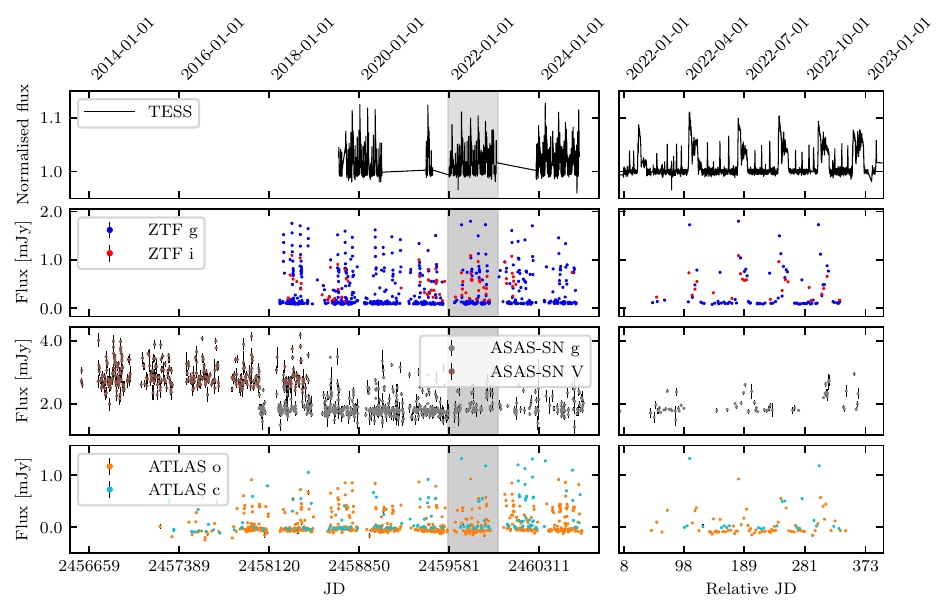}   
    \end{minipage}
 
   \caption{Comparison of TESS and ground-based light curves of KL Dra. The left column displays the full time span of available data, while the right column zooms into a 400-day window starting from JD 2459573, highlighted in grey in the left panels. The top row presents the cleaned and corrected TESS light curves, where black lines indicate 1-hour binned data. The second, third, and fourth rows show cleaned and nightly-binned light curves from ZTF, ASAS-SN, and ATLAS, respectively, in different filters. The high-cadence, continuous monitoring of TESS is evident in the right column, while the long-term but lower-cadence coverage from ground-based surveys is prominent in the left column.}\label{fig:alldata}
\end{figure*}

\subsection{Space-based light curves}\label{subsec:TESSlcs}

Light curves for the pre-selected sources in TPFs are produced by the mission at the Science Processing Operation Center (SPOC) \citep{Jenkins:2016}. The automated pipeline consists of initial calibrations and a process known as Simple Aperture Photometry (SAP) which extracts and corrects the flux from the source by defining masks for the background and the target directly from the TPFs\footnote{TPFs have a nominal square size of 10 pixels.}. Light curves with an additional level of processing known as Presearch Data Conditioning Simple Aperture Photometry (PDCSAP) are also generated, where removal of systematics corrects for the common trends observed in the flux measurements of a given sector through cotrending basis vectors (CBVs) \citep{Smith:2012, Kinemuchi:2012}. We do not use the PDCSAP light curves because despite offering a higher level of systematic correction, they are not optimised for transient activity. Therefore, in most cases the application of CBVs suppresses or distorts the outbursts we aim to study.

Our starting point for the analysis of our TESS data is either the FFIs or the SPOC's TPFs. To obtain the light curves from these images we perform a custom version of the SAP processing steps for the data at each sector. For FFI data we made 11 $\times$ 11 pixel cutouts of the FFI \citep{Brasseur:2019ascl.soft} and on these TPF-size images we define aperture masks around the position of KL Dra, as indicated by the WCS of the image. We define as the background mask the darkest 25 pixels ($\sim$20\% of the TPF) which are obtained by stacking the whole set of exposures in a sector and sorting the pixels by their flux value. Once the masks are defined, we extract the light curves and correct the aperture fluxes by scaling our measurement of the background to the number of pixels in the aperture mask. We created a set of 14 predefined apertures \citep[3-9 pixel each, similar to][]{Feinstein:2019PASP} and for each sector we extract light curves with all of them, selecting the best aperture mask by visually inspecting all light curves. For the extended mission data with high level data products, our light curve extraction process is mostly the same, with the only difference being that we start from the SPOC TPFs. We do not correct for background since the TPF pixel values are already background corrected. To determine the aperture mask, we include the aperture defined by the SPOC in comparison with our predefined masks.

The highly complex TESS background often leaves slow residual trends in the baseline of the SAP-like source light curves, even after the pixel-level background subtraction. To remove this residual background, we fit a low-degree Chebyshev polynomial to the quiescent portions of the source light curve in each sector, masking all high state features. We then normalise the light curve by this baseline, which sets the background level to unity\footnote{TESS flux measurements are given in units of electron counts per seconds.}. Since each TESS sector corresponds to observations performed during two orbits of the spacecraft around the Earth, it is very common to observe patterns that repeat for each orbit due to scattered light from the Earth or the Moon. When these patterns or significant instrument systematics are evident we perform the detrending of the background per orbit instead of per sector. In some cases of highly contaminated light curves, for example when the fine pointing is deactivated during the momentum dumps, we split even further the orbit light curves or reject the contaminated sections. The 20 sec TESS data exhibit significant scatter even in the brightest parts of the KL Dra light curve, making difficult the visual separation of the outburst features and the background. Therefore, during the background detrending process we binned the light curves per sector into 1- or 2-hour intervals in order to identify the region limits for our background fits. 

We implemented an outlier detection and removal algorithm to the background-corrected TESS lightcurves per sector. In particular, we used the Local Outlier Factor (LOF) algorithm \citep{breunig:2000} to flag outliers in both flux and flux error space by comparing the local density of samples with respect to the density of their neighbours. 

The final TESS light curves are presented in the first row of Figure \ref{fig:alldata}. The observations span 1953 days ($\sim$5 years, from 2019-07-18 to 2024-11-21), of which nearly 1000 days correspond to continuous monitoring of KL Dra, with a total of $\sim$2.8 million measurements. The 30-min cadence dataset contains 14,035 data points, equivalent to $\sim$290 days of observations. The 20-sec cadence dataset comprises 2,730,589 data points, covering $\sim$630 days. The 120- and 200-sec datasets together include 49,955 points, corresponding to $\sim$100 days of coverage. Note that the TESS light curve is expressed in flux units normalised to the background level. For the remainder of this work, we vertically shift the TESS light curve by subtracting 1, so that the background level corresponds to zero flux, and flux values represent amplitudes relative to this baseline.

\subsection{Ground-based light curves}\label{subsec:Groundlcs}

We detect and remove anomalous data points in the ZTF, ASAS-SN, and ATLAS light curves based on photometric accuracy. Due to the sparse and irregular sampling of the ground-based data, applying the algorithm used in Subsection \ref{subsec:TESSlcs} for outlier detection is less suitable. For ZTF data, we utilised various metrics and quality flags provided by the light curve extraction pipelines to filter out poor-quality photometric measurements. In the case of ASAS-SN and ATLAS, we retained only measurements with flux errors within 10\% and 5\%, respectively, of the typical flux during KL Dra's high state.

We binned the ZTF, ASAS-SN, and ATLAS light curves on a nightly basis. To achieve this, instead of dividing the time array into equally spaced bins, we applied the Density-Based Spatial Clustering of Applications with Noise (DBSCAN) algorithm \citep{Ester:1996} to the timestamps. With this approach we group measurements within the same night without imposing a fixed bin size. We combined the measurements in each cluster using a weighted mean, where the weights are the inverse of the data point variance, thus giving more weight to measurements with smaller errors. 

The resulting ZTF, ASAS-SN, and ATLAS light curves are presented in the second, third, and fourth rows of Figure~\ref{fig:alldata}. The ground-based dataset spans a total of 4,589 days ($\sim$12.5 years, from 2012-06-01 to 2024-12-25). The cleaned and binned ZTF light curve consists of 694 data points in the ZTF-g filter and 77 in the ZTF-i filter, covering 6.5 years. ZTF-r data was excluded due to an insufficient number of observations. The ASAS-SN dataset includes 600 measurements in the g filter and 354 in the V filter, spanning 12.5 years. The ATLAS dataset, covering 9.4 years, comprises 572 points in the o filter and 173 in the c filter.

To merge the space-based and ground-based datasets into a single light curve, we scale the flux from the ground-based observations to match the normalised flux amplitude of the TESS light curve. We ignore the errors introduced by TESS’s inability to detect the quiescence level of KL Dra, as well as the differences in colour arising from differences in spectral coverage between the filters involved. The scaling process is conducted for each telescope and filter individually and requires downsampling the TESS light curves to match the timestamps of the ground-based data points. Specifically, TESS data is binned into 10-hour windows centred on the times of the ground-based observations. To determine the optimal scaling factor and offset to apply to the ground-based light curves, we minimise the sum of the error-weighted residuals of the flux differences. Due to the continuous sampling of TESS data, there is significant overlap between TESS and the ground-based observations in the ZTF g, ZTF i, ASAS-SN g, ATLAS o, and ATLAS c filters, as shown in Figure \ref{fig:alldata}. However, observations in the ASAS-SN V filter stopped before TESS began collecting data on KL Dra. Therefore, to scale the ASAS-SN V data, we applied the same optimisation process by minimising the residuals against the combination of the other five already-scaled ground-based light curves. The resulting scaling factors were 13.6 and 9.5 for the ZTF g and i bands, 16.3 and 14.6 for the ASAS-SN g and V bands, and 9.5 and 12.8 for the ATLAS o and c bands, respectively. Note that in Figure \ref{fig:alldata}, the flux of KL~Dra measured by ASAS-SN is significantly higher than that reported by ZTF and ATLAS. This is due to the presence of a background galaxy located just 5.7 arcseconds from KL~Dra. Given ASAS‑SN’s relatively large pixel scale (8 arcsecs, about four times larger than those of ZTF and ATLAS), its baseline flux is consequently much brighter. Nevertheless, the flux variations still capture the transient behaviour of the binary, just as TESS does, despite its much larger pixel scale.

\subsection{Outburst activity modelling} \label{subsec:outact}

\begin{figure}
    \centering
    \begin{minipage}{.50\textwidth}
        \centering
        \includegraphics[width=\linewidth]{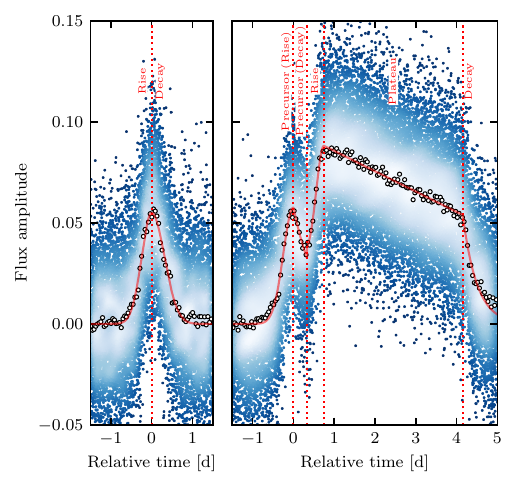}   
    \end{minipage}
 
    \caption{Example of a NO (left) and a SO (right) from TESS data. The blue dots represent the 20 sec full-cadence data, where the colour intensity indicates point density based on a Kernel Density Estimator, with darker shades for lower densities and lighter shades for higher densities. The white circles with black edges show 1-hour error-weighted binned data. The red solid lines correspond to the best-fit models, while the vertical red dotted lines mark the different evolutionary stages of the NO/SO.}\label{fig:morphology}

\end{figure}

TESS observations of KL Dra SOs have previously been reported by \cite{Duffy:2021} and \cite{Pichardo:2021} for a few sectors. Here we report the long term monitoring of the binary with TESS where we have detected multiple NOs and SOs. We fit functions to the NO and SO profiles in order to make a quantitative characterisation of their amplitudes and durations.  

We have dissected each NO and SO into different components. For each NO we distinguish their rise and decay phases. KL Dra exhibits precursors prior to the plateau phase of the SO and rebrightenings after the SO decay. In this work we consider the precursor as part of the SO. Therefore we divide each SO into the precursor, rise to the plateau, plateau phase, and decay from the plateau (see Figure \ref{fig:morphology}). 

For NOs, we model the rise phase as a Gaussian from the noise level to the brightness peak, followed by the decay phase, modelled as another Gaussian from the peak to the TESS noise level as defined in equation \ref{eq:NO}.

\begin{equation}
F_{NO}(t)=\begin{cases} 
F_\text{NO} \ e^{-\frac{(t-t_\text{NO})^2}{2 \ \tau_\text{riseNO}^2}} & t<t_\text{NO}, \\ 
F_\text{NO} \ e^{-\frac{(t-t_\text{NO})^2}{2 \ \tau_\text{decayNO}^2}} & t \geq t_\text{NO}.
\end{cases}
\label{eq:NO}
\end{equation}

Where $t_\text{NO}$ is the time of maximum in the NO light curve, $F_\text{NO}$ its peak flux, $\tau_\text{riseNO}$ is the timescale of the rise and $\tau_\text{decayNO}$ is the timescale of the decay.

We have modelled the components of the SOs and their precursors with the functions given in equation \ref{eq:SO}. 

\begin{equation}
F_{SO}(t) = \begin{cases} 
F_\text{Pre} \ e^{-\frac{(t - t_\text{Pre})^2}{2 \tau_\text{risePre}^2} } & t < t_\text{Pre} \\
F_\text{Pre} \ e^{-\frac{(t - t_\text{Pre})}{ \tau_\text{decayPre}} } & t_\text{Pre} \leq t < t_\text{rise}  \\
F_\text{rise} + \Delta F_\text{rise}  \ e^{-\frac{(t - t_\text{plat})^2}{2 \tau_\text{rise}^2} } & t_\text{rise} \leq t < t_\text{plat} \\
F_\text{plat} + \frac{\Delta F_\text{plat}}{\delta t_\text{plat}} (t - t_\text{plat}) & t_\text{plat} \leq t < t_\text{decay} \\
F_\text{decay} \ e^{-\frac{(t - t_\text{decay})}{\tau_\text{decay}} } & t \geq t_\text{decay}
\end{cases}
\label{eq:SO}
\end{equation}

\begin{figure*}
    \centering
    \begin{minipage}{\textwidth}
        \centering
        \includegraphics[width=\linewidth]{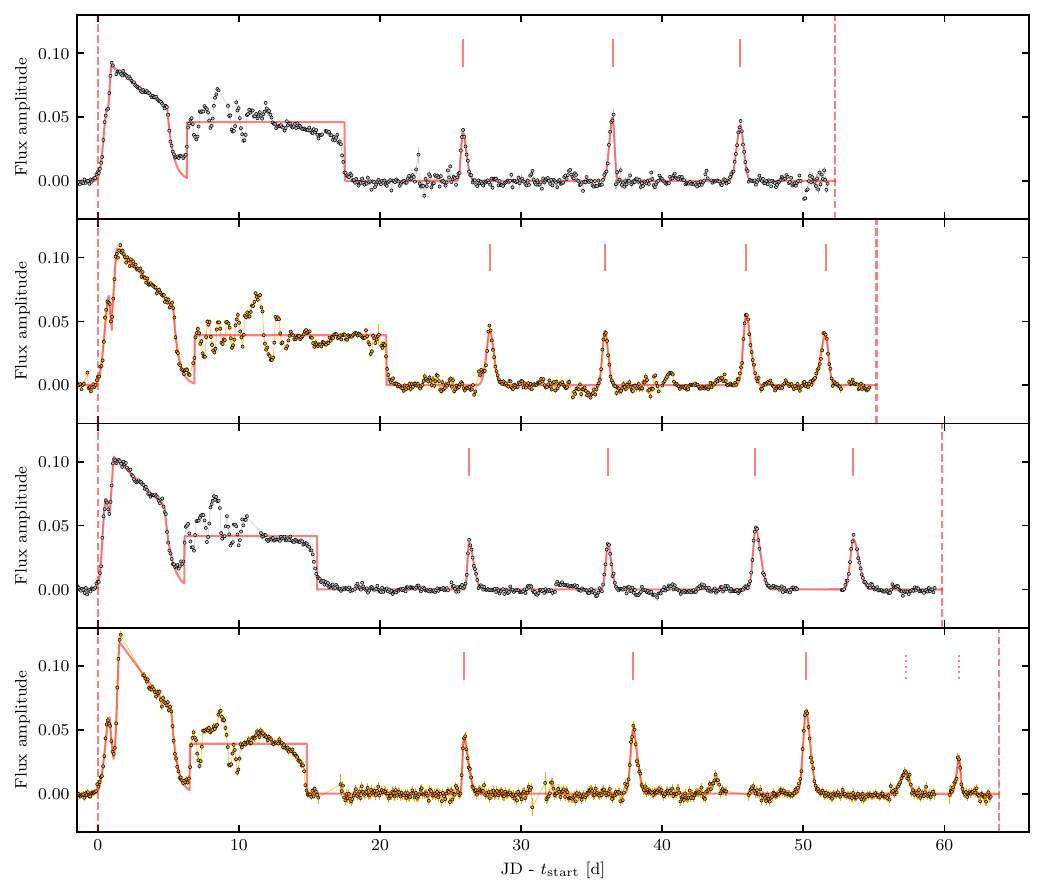}   
    \end{minipage}
    \caption{Sample of KL Dra SCs with complete TESS coverage, shown as a function of time from the onset of each SC. Each row corresponds to a different SC.
    The observed TESS data has been downsampled into 2-hour bins, with the best-fit model shown as a solid red line. Red vertical dashed lines mark the start and end times of each SC, while small solid and dotted red lines indicate the positions of NOs and mini-NOs, respectively. SCs are sorted in ascending order of duration from top to bottom but are not sequential in time. All are displayed on the same time scale for consistency.}\label{fig:SC_selected}
\end{figure*}

Where $t_\text{Pre}$ and $F_\text{Pre}$ are time and flux of the precursor maximum while $\tau_\text{risePre}$ and $\tau_\text{decayPre}$  their rise and decay timescales. Note that instead of modelling the decay component of the precursor with a Gaussian as in the NO decay, we used an exponential decay. We modelled the rise to plateau stage as a Gaussian rise of timescale $\tau_\text{rise}$ with maximum $F_\text{plat}$ at $t_\text{plat}$, starting from $F_\text{rise}$. We modelled the plateau stage as a linear decrease in the brightness where $\Delta F_\text{plat}$ and $\delta t_\text{plat}$ are the amplitude and duration. The final stage is modelled as an exponential decay, parametrised with the time at the start of the decay $t_\text{decay}$, its peak flux $F_\text{decay}$ and timescale $\tau_\text{decay}$. 

We adopted different functional forms for the decay phases of the NO and the precursor, despite accretion models predicting them to be physically the same. We found qualitatively that the SO profile is better reproduced when the precursor decay phase is described by an exponential function rather than a Gaussian, motivating our choice of the former. For the NOs we opted for a Gaussian decay to enable a direct comparison with the modelling approach of the rise phase.

A detailed modelling of KL Dra rebrightenings is beyond the scope of this work.  However, we characterise their onset ($t_\text{rebStart}$), duration ($t_\text{rebEnd}-t_\text{rebStart}$), and flux ($F_\mathrm{reb}$) using a simple box model, defined as  $F_\mathrm{reb}(t) = F_\mathrm{reb}$  for  $t_\text{rebStart} \leq t \leq t_\text{rebEnd}$.

To determine transition times between the background level and the NO/SO phases, we leverage the fitted timescales  $\tau$. Since the NO rise, NO decay, and precursor rise are modelled as Gaussians, their respective $\tau$  values ($\tau_\mathrm{riseNO},\,\tau_\mathrm{decayNO},\,\tau_\mathrm{risePre}$) represent standard deviations. We use the 2.5$\sigma$ limit, which encompasses $\sim99\%$ of a Gaussian distribution, to define the NO onset, NO end, and SO onset. Conversely, as the SO decay follows an exponential model where  $\tau$  represents the time to decay by a factor of  $e$, we define the SO end as $1.5 \tau_\mathrm{decay} $. 

For regions lacking TESS coverage, we use the scaled ground-based data to constrain SO onset and, for well-sampled events, the start and end times of rebrightenings. Depending on data quality, we apply three different template models. For the lowest-quality ground-based SOs, one template constrains the SO start; for intermediate cases, a second template additionally constrains rebrightening start and end times; and for the highest-quality SOs, a third template allows simultaneous fitting of SO onset and full rebrightening duration. These templates are derived from the mean parameters obtained in TESS SO and rebrightening fits. To assess uncertainties in the template fits to ground-based data, we apply them to SO+rebrightening events with simultaneous TESS and ground-based coverage, using the TESS results as reference. This comparison quantifies biases and errors introduced by the template-fitting approach.

\section{Results} \label{sec:res}

\begin{figure*}
    \centering
    \begin{minipage}{\textwidth}
        \centering
        \includegraphics[width=\linewidth]{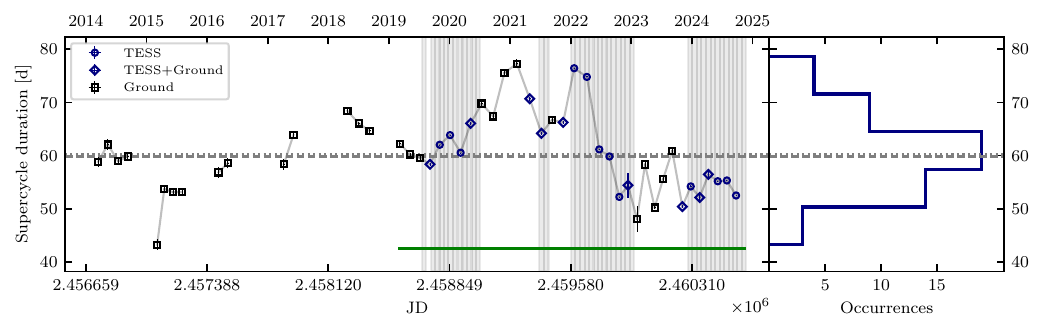}   
    \end{minipage}
   \caption{Time evolution (left) and histogram (right) of KL Dra SC durations. Blue circles indicate SCs with both start and end times constrained by TESS data. Blue diamonds represent cycles where either the start or end was measured using ground-based data (ZTF, ASAS-SN and/or ATLAS), while black squares denote cycles constrained solely by ground-based observations. A grey line connects successive SC measurements. The horizontal green line marks the period over which 34 consecutive SC measurements were obtained. The grey dashed line represents the median duration while the dotted line the mean. Vertical grey shaded regions indicate the time spans covered by individual TESS sectors.}\label{fig:SCdur}
\end{figure*}

\begin{figure}
    \centering
    \includegraphics[width=1\linewidth]{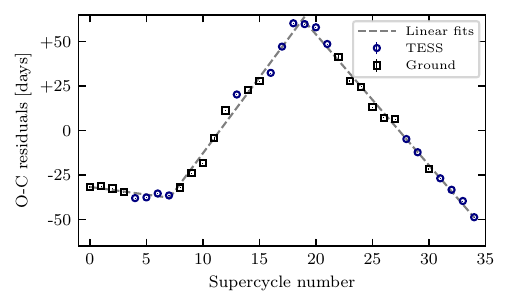}
    \caption{O-C residuals of SC start times relative to a linear ephemeris, as a function of SC number. The analysis includes the continuous sequence of 34 SC measurements from the second half of our dataset, marked by the horizontal green line in Figure~\ref{fig:SCdur}. Blue circles indicate start times derived from TESS observations, while black squares denote those from ground-based data. The grey dashed line represent linear fits to distinct segments of the O-C residuals, highlighting changes in SC timing behaviour.}
    \label{fig:SC_OC}
\end{figure}

The results of our analysis of the decade-long monitoring of KL Dra's outburst behaviour are presented in order of decreasing characteristic timescale. We begin with SCs, followed by rebrightenings, SOs and their individual components, and conclude with NOs. Each section provides a qualitative overview of the feature as revealed by TESS, followed by a quantitative characterisation and statistical analysis using both space- and ground-based datasets.

When multiple measurements of a given quantity are available, we implement a Monte Carlo approach to account for their associated uncertainties. We generate $10^4$ realisations of the dataset, assuming each measurement is drawn from a Gaussian distribution with a mean equal to the measured value and a standard deviation equal to the measured error. For each realisation, we compute the mean, median, and standard deviation as measures of central tendency and dispersion. The final reported values correspond to the averages of these means, medians, and standard deviations across all realisations, with uncertainties in these values given by the standard deviation of the simulated averages. The same Monte Carlo procedure is applied when estimating the Pearson and Spearman rank correlation coefficients. 

This work focuses on variability occurring on timescales of days. While a detailed investigation of shorter-timescale phenomena is beyond the scope of this paper, we note that superhumps are clearly detected during the plateau phases of all TESS SOs beginning with TESS Cycle 4, when the cadence became adequate to resolve them for the first time. In contrast, no superhumps were detected during rebrightenings or NOs. A comprehensive periodicity analysis of the high-cadence TESS observations is in progress and will be presented in Salazar Manzano et al. (in prep).

\subsection{Supercycles} \label{subsec:SCs}

\begin{figure}
    \centering
    \includegraphics[width=1\linewidth]{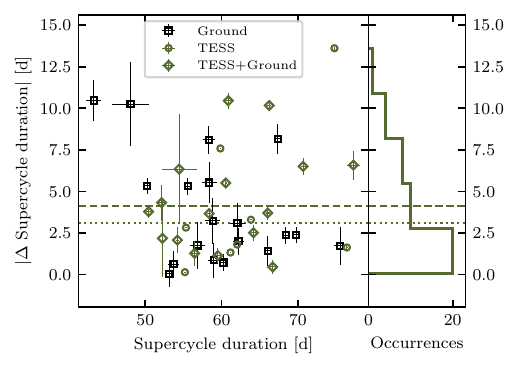}
    \vspace{0.01cm} 
    \includegraphics[width=1\linewidth]{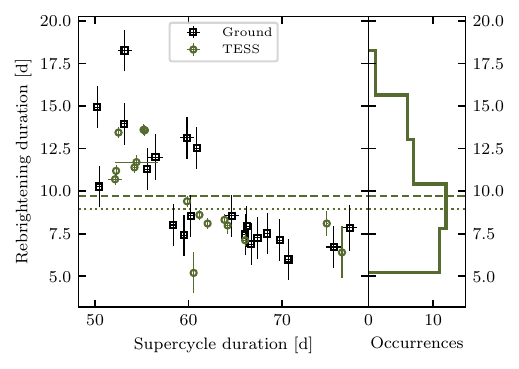}
    \caption{Change in SC duration (top), and rebrightening duration (bottom), as functions of SC duration. The top panel measures the absolute difference between the duration of a given SC (shown on the x-axis) and the immediately following SC. Marker shape and colour indicate whether measurements were obtained from TESS, ground-based data, or a combination of both. The histograms to the right display the distribution of the measured changes, the horizontal dashed line the median, and the horizontal dotted line the mean.}
    \label{fig:combined-figures}
\end{figure}

\begin{figure}
    \centering
    \includegraphics[width=1\linewidth]{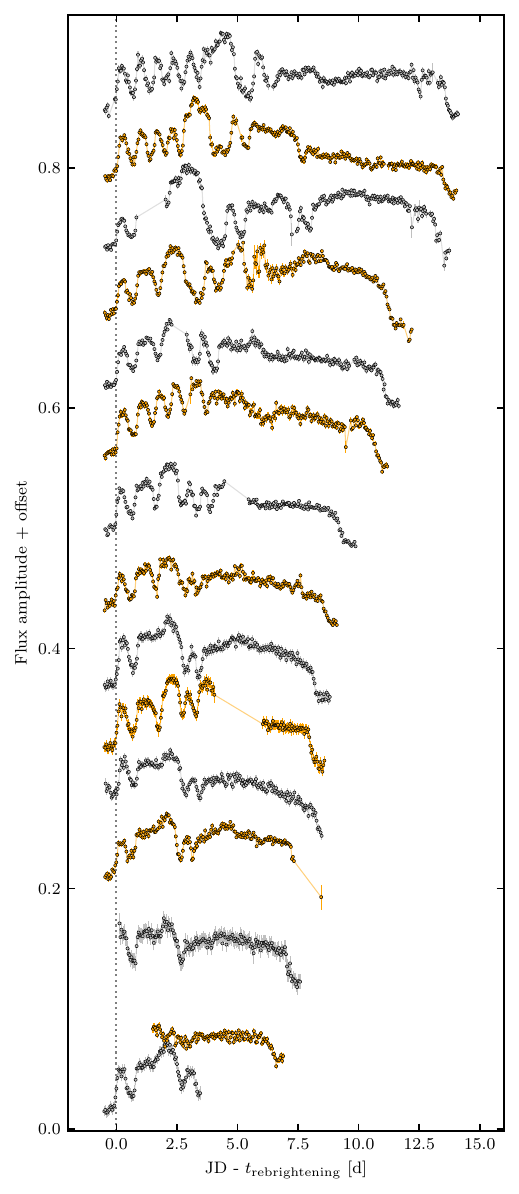}
    \caption{
    KL Dra rebrightenings observed by TESS. The points represent the 1-hour bins, with 2 different colours for visualisation purposes. From bottom to top, the rebrightenings are arranged in order of increasing duration, with a constant vertical offset of 0.06 added for clarity. All rebrightenings are aligned at their onset. Some rebrightenings exhibit data gaps (see text for details).}
    \label{fig:Rsvertical}
\end{figure}

\begin{figure}
    \centering
    \includegraphics[width=1\linewidth]{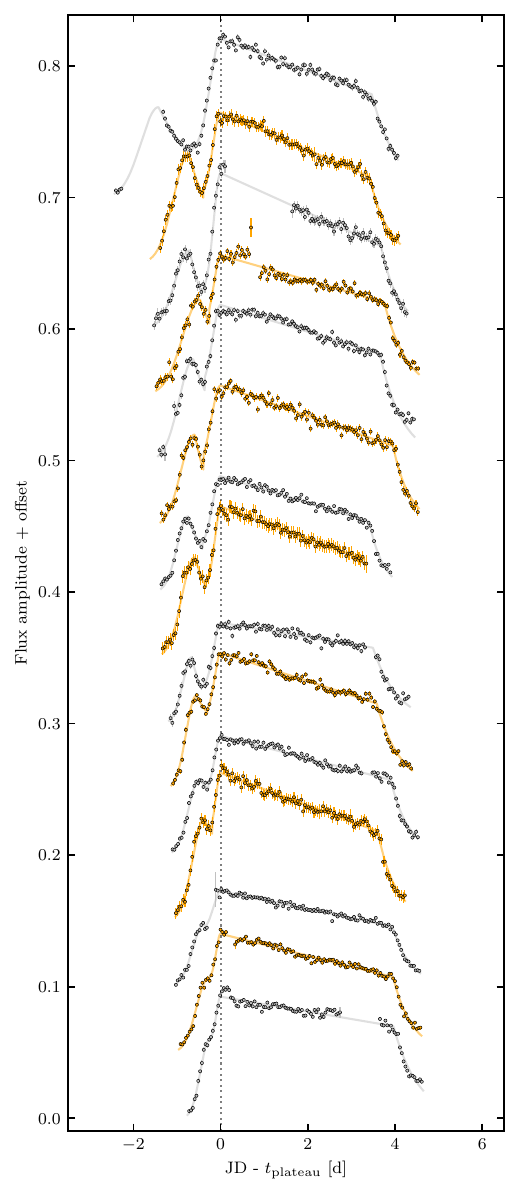}
    \caption{KL Dra SOs observed by TESS. The points represent 1-hour bins, while the solid lines indicate the best-fit models, aligned at the start of the plateau phase, with two different colours used for visual clarity. From bottom to top, the SOs are arranged in increasing order of time between the SO onset and the plateau start, with a constant vertical offset of 0.05 added for clarity.}
    \label{fig:SOsvertical}
\end{figure}

\begin{figure}
    \centering
    \includegraphics[width=1\linewidth]{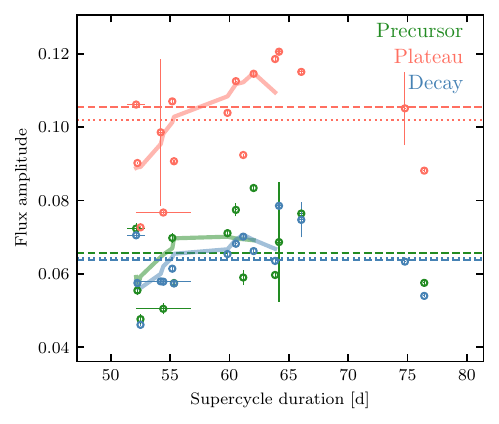}
    \caption{TESS-measured amplitudes of the precursor (green), plateau (pink), and SO decay (blue) as a function of SC duration. The dashed and dotted lines represent, respectively, the median and the mean. Solid lines correspond to a running average computed via convolution, added for visualisation purposes.}
    \label{fig:SOs_amplitudes}
\end{figure}

\begin{figure*}
    \centering
    \begin{minipage}{\textwidth}
        \centering
        \includegraphics[width=\linewidth]{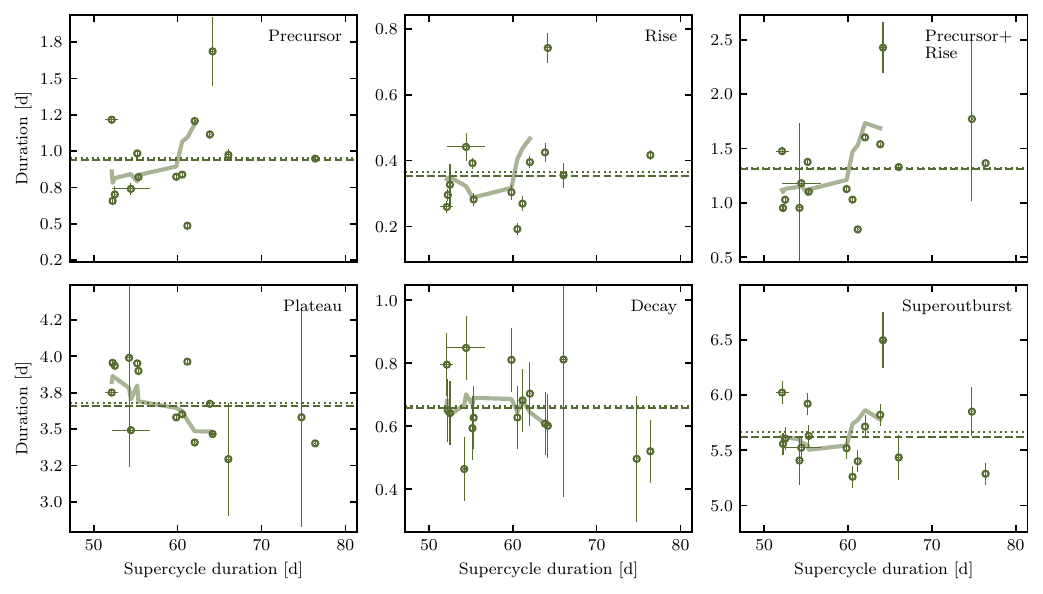}   
    \end{minipage}
    \caption{
    Duration of different SO phases as a function of SC duration. Each panel represents a specific SO phase, including precursor, rise to plateau, plateau, decay from plateau, and total SO duration. The green dashed line indicates the medians and the green dotted line the means, while the solid line represents a running average added for visualisation purposes. }\label{fig:SOvsSC}
\end{figure*}

To determine the SC duration or recurrence time of the SOs, we measured the interval between the onset of consecutive SOs. A total of 49 KL Dra SCs were identified in the combined space- and ground-based light curves. The time span between the first and last identified SC in our dataset is $\sim$10.7 years. Of these, 14 have nearly continuous TESS coverage, and a selection of these is shown in Figure~\ref{fig:SC_selected}. This sample illustrates both the regularity and variability of KL Dra’s SCs. The regularity is evident in the consistent structure of all observed SCs: a SO, immediately followed by a rebrightening phase, and then at least 3 NOs. The variability arises from differences in SC duration, as well as in the amplitudes and durations of individual outburst components within each cycle. The complete dataset of TESS and ground-based SCs, along with their models, is provided in Appendix~\ref{app:SCs}. Note that some gaps in the TESS data (e.g., the plateau gap in the bottom panel of Figure~\ref{fig:SC_selected}) are due to scheduled downlinks, inter-sector gaps, or segments excluded because of high contamination. 

The temporal evolution of SC durations is presented in Figure \ref{fig:SCdur}. The first half of the dataset comprises 15 SC duration measurements, grouped into five sparse segments of 2-4 consecutive SCs constrained solely by ground-based observations. The gaps in this portion of the dataset are mainly a consequence of the annual visibility gaps during which KL Dra is unobservable from Earth (see Figure \ref{fig:alldata}). Over the past $\sim5.7$ years, thanks to the onset of TESS monitoring, we have determined 34 consecutive SC durations. The mean and median SC durations are consistent within uncertainties, with a mean of $60.4\pm0.1$\footnote{Throughout this paper, uncertainties quoted on central values indicate the precision with which the central statistic is estimated, whereas the relevant scatter of the measurements is reported separately as the dispersion of the distribution.} d, with a dispersion of 7.5 d. The distribution of SC durations is approximately Gaussian, though slightly skewed towards shorter durations. The slight skew may reflect an observational bias: shorter SCs are easier to obtain continuous coverage for, whereas longer SCs require extended, uninterrupted monitoring to be confirmed and are more likely to be missed or truncated. The shortest measured SC duration is $43.3\pm0.9$ d, while the longest is $77.3\pm0.8$ d, indicating that KL Dra’s SC duration varies within a range of $33.9\pm1.3$ d.

The time evolution of SC durations over the past $\sim6$ years is not entirely random but it shows some trends. For example, considering the TESS light curves between 2022 and 2023, KL Dra exhibited a steady decrease in SC duration. A similar trend was observed with the ground-based data from 2018 to the end of 2019. Conversely, an increase in SC duration is observed between late 2019 and 2021.

These trends are further supported by the O-C diagram shown in Figure \ref{fig:SC_OC}, which shows the residuals between observed SC start times and a linear ephemeris. The analysis includes only the most recent 34 SCs, for which the combined ground- and space-based coverage guarantees that no SO was missed (see Appendix~\ref{app:SCs}). The best-fit linear ephemeris yields a slope of $61.6\pm0.7$ d and an intercept of $2458582.7\pm15.4$. Negative O-C values correspond to start times that occur later than predicted, whereas positive values indicate an advance. The diagram reveals three distinct regimes: two intervals with decreasing O-C values (but with different slopes), separated by a phase in which the O-C values increase. Taken together, the observed behaviour suggests that, despite the approximately Gaussian distribution of SC durations, consecutive SCs are not fully independent. Instead, they may follow long-term cycles of increasing and decreasing SC duration.

An additional quantitative analysis of the variation in SC duration is presented in the top panel of Figure \ref{fig:combined-figures}. The mean and median changes in SC duration between consecutive SCs are $4.1\pm0.1$ d and $3.1\pm0.2$ d, respectively, with a dispersion of 3.5 d. The distribution of changes is highly asymmetric, spanning from a minimum of $0.05\pm0.8$ d to a maximum observed change of $13.6\pm0.1$ d. Spearman rank and Pearson correlation coefficients indicate no significant correlation between SC duration and its corresponding variation. 

A significant fraction of the SC is spent in quiescence, i.e. when KL Dra is in a low state. As a proxy for this interval, we adopt the time span between the end of a rebrightening (i.e. the end of our box model) and the onset of the subsequent SC. We refer to this interval as post-rebrightening duration. While quiescence technically excludes NOs, which are not constrained in ground-based data, using the post-rebrightening duration allows us to leverage the full ground-based dataset and combine it with the TESS data. The mean post-rebrightening duration is $44.1\pm0.4$ d (consistent with the median), with a dispersion of 9.8 d. The longest observed post-rebrightening phase lasted $63.1\pm1.2$ d, while the shortest was $26.4\pm4.7$ d, yielding a total variation of $36.7\pm4.8$ d. 

\subsection{Rebrightenings} \label{subsec:Res}

All KL Dra rebrightenings observed by TESS are presented in Figure \ref{fig:Rsvertical}, revealing a distinctive pattern. The bottom two rebrightenings in Figure \ref{fig:Rsvertical} are affected by TESS gaps, the lowest is missing the end, and the second-from-lowest is missing the beginning part. The missing start/end times were recovered from ground based data. In general, KL Dra rebrightenings consist of an initial phase with multiple short-lived consecutive outbursts, followed by a relatively steady and longer-lived outburst. A general trend is observed in which the flux of the rebrightening tends to increase over time, as evidenced by the rising amplitudes of consecutive small outbursts. Additionally, a gradual decline in flux is seen from the middle to the end of the rebrightening. 

A quantitative characterisation of rebrightening duration as a function of SC duration is shown in the bottom panel of Figure \ref{fig:combined-figures}. The mean and median rebrightening durations are $9.7\pm0.2$ d and $8.9\pm0.3$ d, respectively, with a dispersion of 3.1 d. The shortest recorded rebrightening lasted $5.2\pm1.2$ d, while the longest was $18.2\pm1.2$ d, implying a variability range of $13.1\pm1.7$ days.

A moderate anticorrelation is observed between the rebrightening duration and the duration of the corresponding SC (bottom panel of Figure \ref{fig:combined-figures}). This relationship follows an approximately linear decline for shorter SCs up to $\sim65$ d, beyond which it appears to flatten towards the longer end of the SC distribution. We obtained a Spearman coefficient of $-0.74 \pm 0.05$ with a p-value of $2.4 \times 10^{-6}$, and a Pearson coefficient of $-0.69 \pm 0.04$ with a p-value of $1.1  \times 10^{-5}$, confirming a moderate negative correlation. The rebrightening duration distribution shows more dispersion for low SC durations. 

\subsection{Superoutbursts} \label{subsec:SOs}

\begin{figure}
    \centering
    \includegraphics[width=1\linewidth]{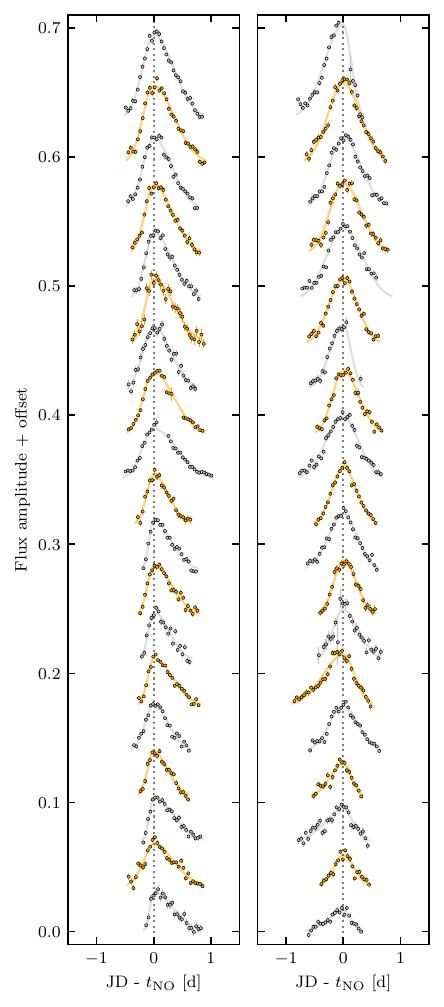}
    \caption{Sample of KL Dra NOs observed by TESS. Points represent 1-hour binned data, and solid lines indicate best-fit models. Two distinct colours are used for clarity. The left panel shows NOs with asymmetry parameters $f_r$ less than 0.4, while the right panel shows those with $f_r$ greater than 0.4. All outbursts are aligned at their peak amplitude, sorted by increasing amplitude, but they are not sequential in time. A constant vertical offset of 0.035 is applied to each NO to aid visualisation.}
    \label{fig:NOs_selected}
\end{figure}

\begin{figure*}
    \centering
    \begin{minipage}{\textwidth}
        \centering
        \includegraphics[width=\linewidth]{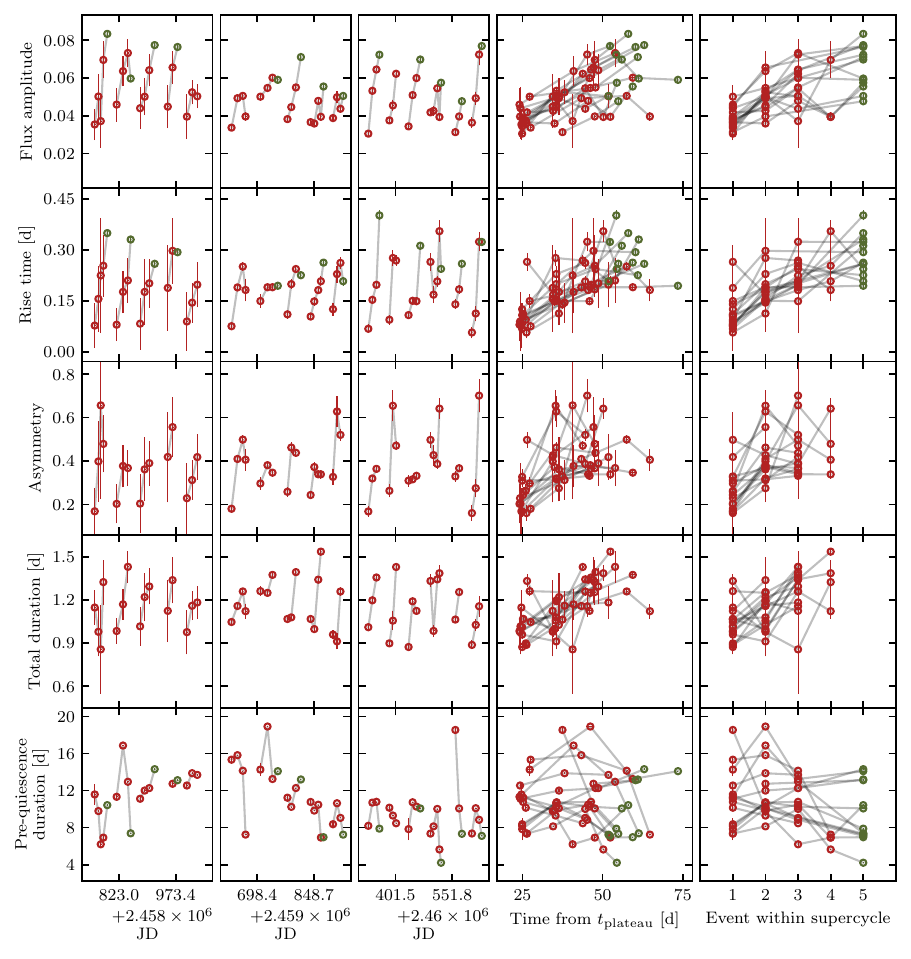}   
    \end{minipage}
   \caption{Time evolution of NO characteristics: amplitude, rise-to-plateau timescale, asymmetry parameter $f_r$, total duration, and pre-quiescence duration. Red points correspond to NOs, and green points to measurements based on precursors (included for amplitude, rise timescale, and quiescence comparisons). Mini-NOs (amplitude $< 0.03$) are excluded from this figure. NOs within the same SC are connected by lines. The first three columns show the absolute time evolution of each characteristic and share a common x-axis scale. The fourth column presents the timing of each NO relative to the onset of the plateau in the corresponding SC. The fifth column shows the order of  occurrence of NOs within each SC, treating precursors as the fifth event. }\label{fig:NOevol}
\end{figure*}

\begin{figure}
    \centering
    \includegraphics[width=1\linewidth]{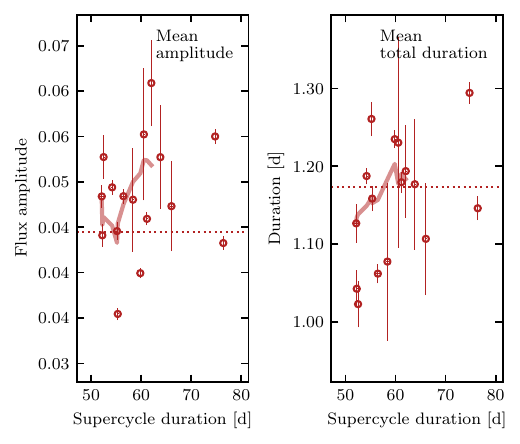}
    \caption{NO properties averaged within each SC as a function of SC duration. The left panel shows the mean NO amplitude, and the right panel shows the mean total duration. Mini-NOs (amplitude $< 0.03$) and precursors are excluded. The dotted line indicates the overall mean, while the solid line shows a running average included for visualisation purposes.}
    \label{fig:NOmedian}
\end{figure}

The light curves and corresponding fits for all KL Dra SOs observed with TESS are shown in Figure \ref{fig:SOsvertical}. The two SOs in the bottom of Figure~\ref{fig:SOsvertical} may appear, at first glance, to lack a precursor. However, we obtain a qualitatively better fit to the data for these cases by retaining a precursor term in Equation~\ref{eq:SO} but removing its decay phase, than by fitting an SO with no precursor (i.e., a Gaussian rise to the plateau only). Taken together, the fits suggest that a precursor is likely present in every detected SO, though its duration and amplitude vary substantially.

The relationship between the precursor, plateau, and decay amplitudes of SOs observed by TESS and the corresponding SC duration is illustrated in Figure \ref{fig:SOs_amplitudes}. The precursor phase has a mean amplitude of $6.6 \times 10^{-2}$, with a 1-sigma uncertainty of $0.1 \times 10^{-2}$, and the dispersion of the measured values is $1.2 \times 10^{-2}$. For the plateau phase, the mean and median amplitudes are $(10.2\pm0.1) \times 10^{-2}$ and $(10.6\pm0.1) \times 10^{-2}$, respectively, with a dispersion of 1.6$\times 10^{-2}$. For the decay phase, the mean amplitude is $(6.56\pm0.03) \times 10^{-2}$, with a dispersion of 0.91 $\times 10^{-2}$. The amplitude of the precursor phase is consistent with the amplitude of the decay phase. In all three SO phases, the lowest recorded amplitude is $\sim60\%$ of the maximum recorded value.

No conclusive correlation was found between the SC duration and the amplitudes of the precursor, plateau, or decay phases. We obtained Spearman rank and Pearson correlation coefficients of $\sim0.3$, with p-values of $\sim1\times10^{-1}$. However, as evident from the running averages in Figure \ref{fig:SOs_amplitudes}, a clearer trend emerges for SCs shorter than $\sim70$ d. When restricting the sample to these shorter SCs, the correlation strengthens. Particularly for the plateau and decay amplitudes, which exhibit a moderate positive correlation of $\sim0.7$ with p-values of $\sim1\times10^{-2}$. Due to the limited number of TESS observations at the long-duration end of the SC distribution, we cannot robustly assess whether similar correlations persist for longer SCs.

The measured durations of the different SO phases as a function of the SC duration are presented in Figure \ref{fig:SOvsSC}. The mean duration of the precursor phase is $0.95\pm0.02$ d, with a dispersion of 0.29 d. For the rise-to-plateau phase, the mean duration is $0.36\pm0.01$ d, with a dispersion of 0.13 d, while the combined duration of the precursor and rise phases is $1.32\pm0.06$ d, with a dispersion of 0.47 d. The plateau stage has a mean duration of $3.68\pm0.06$ d, with a dispersion of  0.34 d, whereas the decay phase has a mean duration of $0.66\pm0.04$ d, with a dispersion of 0.18 d. The total duration of the SO event, from onset to the end of the decay phase, has a mean duration of $5.67\pm0.03$ d and a dispersion of 0.34 d.

We found no conclusive correlations between the durations of the SO phases or total SO duration and the SC duration. However, as with the SO amplitudes, the correlation coefficients and p-values tend to strengthen when restricting the sample to SCs shorter than 70 d. The strongest indication arises for the plateau duration, which shows a correlation coefficient of $\sim-0.5$, though the associated p-value of $\sim1\times10^{-1}$ remains statistically insignificant.

\subsection{Normal outbursts} \label{subsec:NOs}

A total of 70 NOs were identified in the TESS data. We determine the asymmetry $f_r$ of a NO by computing the ratio of its rise phase duration to its total duration, meaning that low $f_r$ NOs have short rise phases compared to their decays. In Figure \ref{fig:NOs_selected} we show a sample of 38 NOs for which the relative errors in asymmetry and amplitude are below 15$\%$. The figure reveals that NOs typically have short durations ($\sim1$ d), and that low $f_r$ events are highly abundant, while a broad range of amplitudes is observed across both small and intermediate $f_r$ values.

NOs were detected in all SCs with continuous TESS coverage from the end of the rebrightening phase through the end of each SC. Of the 70 NOs, 61 belong to these well-sampled SCs, and their temporal evolution is illustrated in Figure \ref{fig:NOevol}. Some NOs exhibit significantly lower amplitudes compared to most NO events within the same SC. This motivates our distinction between the traditional NO and ``mini-NO'', with the latter defined as events having a normalised amplitude smaller than 0.03. In our dataset, 11 outbursts fall into this mini-NO category.

The NO amplitude (excluding mini-NOs) shows an increasing trend within each SC, as seen in the first row of Figure \ref{fig:NOevol}. To provide additional insight into this trend, we also included the amplitude of precursors (green points). This increasing trend becomes more evident when the NO amplitude is shown as a function of time relative to the onset of the plateau phase within each SC\footnote{We use as a reference the onset of the plateau rather than the start of the SC, as the former is generally determined with lower uncertainty.} (fourth column in Figure \ref{fig:NOevol}). A similar positive correlation is observed when ordering the NOs by their sequence within the SC (top-right panel of Figure \ref{fig:NOevol}). Excluding both precursors and mini-NO, we find Spearman and Pearson correlation coefficients of $\sim0.5$ with p-values of $\sim10^{-3}$, for a relationship between NO amplitude and time from the plateau onset. When the precursors are included (excluding only mini-NOs), the correlation strengthens, with coefficients of $\sim0.6$ and p-values of $\sim10^{-7}$. If mini-NOs are included, the correlation coefficient is $\sim$0.45 with a p-value of $\sim10^{-4}$. This moderate positive correlation is further supported when examining amplitude as a function of event order within each SC, where we find coefficients of $\sim0.65$ with $\sim10^{-7}$ p-values.

The fourth column of Figure \ref{fig:NOevol} indicates that NOs (excluding mini-NOs) do not occur uniformly throughout the quiescent phase following the rebrightening. Instead, their timing follows a relatively consistent pattern: the first NO typically occurs $\sim25$ d after the plateau onset, the second near $\sim35$ d, and the third, though more scattered, near $\sim45$ d. Most SCs display three NOs with amplitude larger than 0.03, with a fourth often observed (see fifth column of Figure \ref{fig:NOevol}). Although mini-NOs are expected to be produced by the same mechanism as NOs, their timing does not correlate with the time elapsed since the plateau onset. 

In addition to amplitude, both the rise timescale and total duration of NOs tend to increase over the course of a SC (see second and fourth rows of Figure \ref{fig:NOevol}). The rise timescale trend is evident across NOs, mini-NOs (not shown), and precursors, and is supported by moderately positive correlation coefficients of $\sim0.62$ ($\sim10^{-6}$ p-value) with respect to the time from the plateau onset, and $\sim0.67$ ($\sim10^{-8}$ p-value) with respect to the NO order. For NOs only (excluding mini-NOs), the correlation between the total duration and the time from the plateau onset\footnote{Precursors are excluded from this analysis since their decay is modelled as an exponential rather than a Gaussian, as is done for NOs.} is $\sim0.61$ (p-value $\sim10^{-5}$), and with respect to NO order is $\sim0.56$ (p-value $\sim10^{-4}$). The trend in total duration is not followed by mini-NOs, which tend to be shorter in duration. 

The evolution of the asymmetry factor $f_r$ reveals a nonlinear trend within the SC\footnote{Asymmetry is not reported for precursors, as their decays are modelled with an exponential rather than a Gaussian profile.}. As shown in Figure \ref{fig:NOevol}, the first NOs in each SC tend to have low $f_r$ values (i.e., short rise phases relative to long decays), while the second NO usually exhibits a higher $f_r$. Subsequent NOs typically maintain or slightly reduce this value, resulting in a peak $f_r$ at the second NO, a pattern also seen when including the mini-NOs. Excluding mini-NOs, we find Pearson and Spearman correlation coefficients of $\sim0.44$ (p-value $\sim10^{-3}$) between the asymmetry and the time from the plateau onset. When considered as a function of occurrence order, the correlation is $\sim0.48$ (p-value $\sim10^{-4}$).

We also investigated possible trends for the quiescence period between NOs by measuring the time interval between their peak amplitudes (excluding mini-NOs). For the first NO, the quiescence is defined from the end time of the rebrightening phase. We include the interval from the peak of the last NO to the peak of the precursor as green dots. As shown in the bottom row of Figure \ref{fig:NOevol}, some SC exhibit a steady decrease in quiescence duration, others show a gradual increase, and other a relatively stable quiescence duration. This results in no correlation when evaluated over the entire sample. 

Finally, to explore potential correlations between NO properties and SC duration, we compute the mean NO amplitude and mean total duration within each SC. Figure \ref{fig:NOmedian} displays these mean values (excluding mini-NOs and precursors) as a function of SC duration for SCs with complete quiescence phase coverage. While the running average for SCs shorter than 70 d may hint at a positive trend, Pearson and Spearman coefficients reveal no significant correlation between SC duration and either NO amplitude or duration. 

\section{Discussion} \label{sec:dis}

\subsection{Comparison with previous KL Dra studies}

This work presents the first comprehensive study of KL Dra's outbursting properties that simultaneously leverages extensive long-term monitoring and dense, high-cadence sampling. Earlier investigations either tracked KL Dra's SOs over many cycles with ground-based telescopes or used a handful of TESS sectors. We find a mean SC length of 60.4~$\pm$~0.1 d, consistent with the $\sim$~60~d recurrence times first reported by \citet{Ramsay:2010} and shorter than the 65 d reported by \cite{2025Kojiguchi}. By employing a direct, event-by-event measurement of each SO, we improved upon the estimate of 60~$\pm$~3~d from \citet{Duffy:2021}, which was based on period folding techniques. Our shortest observed SC of 43.3~$\pm$~0.9~d closely matches the 44~d minimum noted by \citet{Ramsay:2012_2}, while our longest SC of 77.3~$\pm$~0.8~d exceeds their reported maximum of $\sim$~68~d, revealing a broader variation range.

\cite{Pichardo:2021} recognised that SO durations were previously overestimated because low-cadence ground-based monitoring could not clearly separate SOs from rebrightenings. Consequently, the 10 d reported by \cite{Ramsay:2012_2} and the more recent $10\pm0.7$ d from \cite{Duffy:2021} measure the interval from SO onset to the end of the rebrightening. In contrast, our dataset resolves these phases separately: we find a mean SO-onset-to-rebrightening-end interval of $16.3\pm0.4$ d, while the intrinsic SO itself lasts $5.67\pm0.03$ d on average. Whereas \cite{Ramsay:2012_2} quoted a 9-17 d range for the SO-onset-to-rebrightening-end interval, we measure a broader span from $11.2\pm1.2$ d to $24.3\pm1.1$ d (a variability range of $13.1\pm1.7$ d). Although \cite{Ramsay:2012_2} suggested that SO peak brightness correlates with SC length, our data seems to suggest this trend only for plateaus and decay amplitudes in SCs shorter than $70$ d (Figure \ref{fig:SOs_amplitudes}). Finally, whereas \cite{Ramsay:2010} reported the time between SO onset and plateau start as $<2$ d, our measurement is $1.32\pm0.06$ d.

Although the distinction between SOs and rebrightenings was not clearly established, both \cite{Ramsay:2012_2} and \cite{Duffy:2021} noted a “dip” feature occurring about $5$ d after SO onset, which they hypothesised could be due to rebrightenings.  \cite{Ramsay:2012_2} reported the dip lasting 2-3 d, which in our framework corresponds to the time between SO decay start and rebrightening start. We measure this with a mean of $1.44\pm0.09$ d and a dispersion of $\sim0.4$ d.  \cite{Pichardo:2021} first recognised for KL~Dra that the duration and amplitude of rebrightenings increase over time.  Our more complete dataset refines this picture, showing the trend continues until roughly the midpoint of the rebrightening, after which a significantly longer outburst decays into the sharp drop that marks its end (see Figure~\ref{fig:Rsvertical}).

As noted earlier, NOs were hinted at in ground-based analyses \citep{Ramsay:2010, Ramsay:2012, Ramsay:2012_2}, but not securely confirmed until TESS observations \citep{Duffy:2021, Pichardo:2021}. \cite{Duffy:2021} reported KL Dra exhibiting 3-4 NOs lasting ~1 d each, consistent with our findings. We measure a mean NO duration of 1.17~$\pm$~0.01 d and identify mini-NOs with a mean duration of 0.87~$\pm$~0.08 d. \cite{Duffy:2021} also noted that shorter SCs lead to shorter intervals between NOs. Our data suggest the strongest correlation is between the SC duration and the maximum interval between NOs within each SC. This arises from the strong correlation between post-rebrightening duration and SC duration. Since longer SCs imply longer post-rebrightening durations, and given the number of NOs per SC remains roughly constant while their total duration is independent of SC duration (Figure~\ref{fig:NOmedian}), the maximum NO separation tends to increase with SC length.

The only dedicated AM CVn modelling to date is by \cite{Kotko:2012A&A} who modelled, among other systems, KL Dra assuming that mass transfer is enhanced during outburst as a result of irradiation of the secondary. At the time, NOs had not been confirmed, so their modelling assumed SO-only light curves (though they did note NOs could have been missed). Additionally, their assumed SO duration was ~14 d, influenced by the ground-based observational issues discussed above. Therefore, our KL Dra measurements are better compared to their modelling of PTF 1J0719+4858, the only AM CVn with confidently detected NOs at that time. Their model predicts features such as precursors and increasing NO amplitudes within a SC, which we observe. Their ‘Z Cam’ AM CVn model reproduces rebrightenings, although not all modelled SCs exhibit them, contrary to what we observe. Their SO precursors do not exhibit the morphological diversity we observe in KL Dra, a difference that arises naturally from the contrast between idealised modelling and the complexity of real systems. A precursor is simply a NO that fails to decay to quiescence and instead triggers the SO once the disc surface density near the outer edge reaches its critical value, the subsequent enhancement in mass transfer suppresses the cooling front and initiates the SO. Because the simulated disc evolve towards a steady, repeating limit cycle, the initial conditions are forgotten, producing nearly identical precursors in each SC. In real systems, however, the mass-transfer rate is not constant, and stochastic fluctuations or non-axisymmetric disc structures (e.g., spiral waves) introduce variability that gives rise to the diversity of precursor profiles we detect. The same reasoning likely accounts for the greater regularity of the simulated light curves relative to the complex and cycle-to-cycle variability observed across the different components of KL~Dra's outburst activity. A dedicated modelling effort incorporating our KL Dra light-curve results is currently underway and will be presented separately.

\subsection{Comparison with previous dwarf novae constraints}

Light‐curve characterisation with cadence and baseline comparable to our study has only been performed for two SU UMa dwarf novae in the Kepler field.  While dwarf novae differ from AM CVns in having lower ionisation temperatures due to their abundance of hydrogen and larger discs, the same underlying thermal instability mechanism is believed to operate.  In particular, Kepler monitored V344 Lyrae and V1504 Cygni with virtually no gaps, revealing SOs of $\sim10$-$15$ d duration and recurrence times of $\sim110$ d, for a total of $\sim10$ SCs detected \citep{Cannizzo:2012Kepler, Osaki:2014PASJ}. These studies also found $\sim10$ NOs per SC, each lasting $\sim2$-$4$ d.

\subsubsection{Precursors and number of outbursts}

The presence of SO precursors was not obvious in the first Kepler light‐curve studies of V344 Lyrae.  \cite{Cannizzo:2010ApJ} and \cite{Cannizzo:2012Kepler} treated them as “shoulders” on the SO because of the shallow dips observed.  The expanded Kepler dataset analysed by \cite{Osaki:2014PASJ} revealed a more diverse range of precursor behaviour, somewhat similar to what we detect in KL Dra’s rebrightenings from TESS (see Figure \ref{fig:SOsvertical}, and discussion in the previous section). \cite{Osaki:2014PASJ} reported a new type of precursor that is widely separated in time from the following SO rise. However, we do not observe this widely separated precursor in our KL Dra TESS data, despite TESS having captured multiple SOs.

A significant difference of KL Dra with respect to the Kepler SU UMa stars is the far smaller number of NOs per SC (3-4 vs $\sim$10). Modelling of these dwarf novae remains inconclusive given the complex interplay among all parameters. The disc’s inner radius ($r_{\mathrm{inner}}$), which can vary due to truncation by the white dwarf magnetic field, can modulate the number of NOs. Increasing the outer disc radius ($r_{\mathrm{outer}}$) raises the count of NOs \citep{Cannizzo:2010ApJ}, consistent with AM CVn systems’ smaller discs, but at the cost of altering the SO profile. Additional parameters that influence the number of NOs (with concomitant changes to other light‐curve properties) include the hot‐ and cold-state viscosity parameter ($\alpha_{\mathrm{hot}}$ and $\alpha_{\mathrm{cold}}$), the mass‐transfer rate (lower $\dot M_{\mathrm{tr}}$ yields less frequent NOs), and the mass ratio $q$ (which sets the disc's tidal truncation radius and hence its extent)  \citep[e.g.][]{Jordan:2024A&A}.

\subsubsection{Outside-in and inside-out outbursts}

Previous works suggest the asymmetry of a NO ($f_r$) provides an indication of where the disc first reaches the critical surface density that triggers the heating wave \citep{Smak:1984AcA,Cannizzo:1986ApJ}. Outside-in outbursts start at the outer disc edge and drive the heating wave inward, producing highly asymmetric NOs with fast rises ($f_r<0.5$) while inside-out outbursts ignite near the white dwarf and propagate outward more slowly, producing more symmetric profiles ($f_r\simeq0.5$). However, we should also note that $f_r$ is not a unique diagnostic of ignition radius: inner-disc truncation (e.g., by a white dwarf magnetic field) can make inside-out bursts appear asymmetric, and in helium-dominated discs the cooling front is slower than in solar-composition discs, making even inside-out bursts more asymmetrical in simulations. The true connection between the observed NO asymmetry and the underlying accretion disc physics remains unclear. In the remainder of this section, we explore possible interpretations of our KL Dra constraints.

\cite{Cannizzo:2012Kepler} found that in both V1504 Cyg and V344 Lyr the NO asymmetry remains relatively constant at $f_r\sim0.2$-$0.4$ within each SC. This contrasts with what we found for KL Dra where the asymmetry parameter changes. For KL Dra the first NO within a SC is asymmetric ($f_r\sim0.2$), while the second or third NOs (third row in Figure~\ref{fig:NOevol}) are usually much more symmetric ($f_r\sim0.4-0.6$). This suggests that for the dwarf novae measured by Kepler \citep{Cannizzo:2012Kepler}, most of the NOs are consistent with outside-in outbursts, but in KL Dra, the first NOs in the SC are in agreement with outside-in outburst expectations, while  the rest of the NOs in the SC are consistent with inside-out type outbursts. Additionally, inside-out outbursts of V344 Lyr have been found to occur after unusually long quiescent periods \citep{Jordan:2024A&A}, which is not the case for KL Dra. 

The three leftmost panels in the bottom row of Figure \ref{fig:NOevol} (corresponding to different TESS windows) show, respectively, an increasing, decreasing, and approximately constant pre-quiescence interval. This pattern closely matches the secular changes in SC  length seen in Figure \ref{fig:SCdur}: when SCs are longer, the mean interval between NOs is also longer, consistent with the nearly fixed number of 3-4 NOs per SC. In the DIM, the recurrence time for inside-out outbursts is set by the viscous diffusion time and is therefore expected to be roughly constant within a SC, contrary to what we observe. For outside-in outbursts, by contrast, the recurrence time shortens as the mass-transfer rate increases. Taken by itself, the observed co-variation of longer SCs with longer inter-NO intervals would thus favour outside-in ignition. 

Within the DIM, the first NO after the SO+rebrightening stage is expected to ignite closer to the inner disc edge and therefore be relatively symmetric. As mass accumulates over successive NOs, the ignition radius should migrate outward towards the disc outer edge, yielding progressively more asymmetric profiles. From the observed inversion in KL Dra (an asymmetric first NO followed by more symmetric subsequent NOs) we speculate that the system’s mass-transfer rate is close to the value at which outside-in outbursts can be triggered. However, we emphasise that the observed inversion remains puzzling. Under this interpretation, not only the first NO but all subsequent NOs should be of the outside-in type. The fact that the second and later NOs appear instead to be inside-out would require the mass transfer rate to decrease during this phase of the SC. Note also that with modest irradiation, the critical $\dot M$ threshold for outside-in outburst can be reached. 


\subsubsection{Outburst duration and amplitudes}

The observations of V344 Lyrae showed that both the duration and amplitude of NOs reach a maximum roughly midway between SOs \citep{Cannizzo:2010ApJ}.  In contrast, our KL Dra light curves (excluding mini‐NOs) exhibit steadily increasing NO duration and amplitude throughout each SC, with the precursor also following the amplitude pattern (see first and fourth rows in Figure \ref{fig:NOevol}).  The growth in NO duration is commonly attributed to the accumulating disc mass after successive outbursts \citep{Cannizzo:2012Kepler,2005Osaki}. The NO dwarf novae behaviour resembles instead the trend seen in KL Dra's rebrightenings discussed earlier, but on much shorter timescales as evidenced by the absence of a quiescent stage in the rebrightenings.  The 2D simulations of V1504 Cyg by \cite{Jordan:2024A&A} differ from the V344 Lyr models of \cite{Cannizzo:2010ApJ}, predicting that all but the first NO in a SC should increase in amplitude.  They explain the first NO’s higher amplitude by the disc’s eccentric configuration following the prior SO.  In our KL Dra monitoring, however, we do not observe a boosted first NO amplitude (see first row in Figure \ref{fig:NOevol}). 

Mini-NOs were previously reported in Kepler light curves of V1504 Cyg \citep{Osaki:2014PASJ}  and we find that KL Dra’s mini-NOs likewise occur randomly throughout the SC.  \cite{Osaki:2014PASJ} proposed that these events might stem from a third, stable branch in the disc‐instability thermal‐equilibrium curve, intermediate between the standard hot and cold branches.  That mechanism, however, implies mini-NOs should appear only at the end of the SC, when tidal dissipation peaks, which conflicts with their random occurrence in both V1504 Cyg and now in KL Dra. By contrast, the standard DIM predicts small amplitude NOs, and they are seen in many simulations. In the DIM, mini-NOs occur when the heating front fails to reach the outer disc edge. Alternatively, a temporary drop in the mass transfer rate can also produce small amplitude NOs: the next outburst begins after a diffusion time, but there is insufficient mass in the outer disc for the heating front to propagate far.     

In V1504 Cyg and V344 Lyr, the quiescent duration between NOs increases until reaching a maximum near half of the SC, after that the duration of the interval decreases \citep{Cannizzo:2012Kepler}.  KL Dra shows no such uniform pattern: some SCs exhibit steadily lengthening quiescent gaps, while others steadily shorten (fifth row of Figure \ref{fig:NOevol}).  The canonical thermal-tidal instability model predicts a monotonic increase in quiescent duration as the disc's outer radius expands after each NO, requiring progressively larger disc masses to reach the critical surface density \citep{Osaki:2005PJAB}, a behaviour we do not observe in KL Dra. In principle, simulations with non-constant mass-transfer rates can reproduce quiescent durations without a clear trend \citep{Jordan:2024A&A}. Additionally, the DIM predicts a correlation between NO amplitude and the interval between NO outbursts (the Kukarkin-Parenago relation) documented across dwarf novae. In KL Dra, however, we find no clear correlation between the pre-quiescence time and the NO flux amplitude (see top and bottom rows of Figure~\ref{fig:NOevol}), strengthening the case that the canonical DIM requires additional ingredients.

Expanding on the DIM perspective, we stress that He discs are highly sensitive: even small changes in metallicity Z can alter stability and the resulting light curve, whereas H discs are comparatively less sensitive. The reason is that, in the temperature range between the first and second ionisation of helium, C, N, and O become highly ionised. Despite their low abundances (e.g., Z=0.02-0.04), they significantly affect the opacity, shifting the critical temperatures and critical surface densities at which the instability sets in. Therefore, in helium discs, even with the same viscosity, mass-transfer rate, and disc radii, modest differences in Z can produce measurably different light curves \citep{Kotko:2012A&A}.

\subsection{Supercycle constraints}

Systematic investigations of recurrence times in AM~CVn systems beyond KL Dra are scarce, and even the rich Kepler datasets for SU UMa dwarf novae have not been analysed for long-term SC evolution. To date, most disc-instability models aim only to reproduce the broad features of one or two SCs, leaving the origins and evolution of recurrence intervals relatively unexplored. This limitation reflects the data landscape until now, as with our combined long- and short-cadence dataset, we can now explore SC behaviour across all timescales.  Indeed, we found a close relationship between SC duration and various outburst properties, such as rebrightening length and SO amplitude, laying the groundwork for more comprehensive disc-instability modelling. For example, it remains to be modelled how mass transfer modulation on timescales longer than a single SC influences changes between consecutive SCs. Our comprehensive multi-SC dataset will allow us to test this in future work and to investigate the origin of a possible ``hypercycle'' enhancement of the mass transfer rate.  

The few studies of SC evolution have focused on SU UMa stars using ground‐based data. \cite{Vogt:1980A&A} was the first to identify variations in recurrence times and to detect repeating periods (cycles) in O-C residual diagrams. In KL Dra, we detect three characteristic cycles (Figure \ref{fig:SC_OC}) from a continuous sample of 34 timed SCs spanning about six years. Traditionally, SC variations have been ascribed to changes in the donor’s mass‐transfer rate \citep{Vogt:1980A&A}. Ultra long outbursts in the AM~CVns SDSS J080710+485259, SDSS J113732+405458 and ASASSN-21au \citep{2020RS, RiveraSandoval:2021, Wong:2021RNAAS, RiveraSandoval:2022} confirm episodes entirely due to mass transfer variations in these systems, although it remains to be seen whether this behaviour applies across all orbital periods. \cite{Simon:2004BaltA} further proposed that fluctuations in the quiescent‐state viscosity, $\alpha_{\rm cold}$, or rapid angular‐momentum removal via magnetic or tidal torques could also drive recurrence time evolution.  Whether these mechanisms apply in AM CVn systems with their more compact scales, helium‐dominated discs, and different donor types remains an open question.

\section{Summary / Conclusions} \label{sec:concl}

Accretion in compact binaries is a high-dimensional problem traditionally framed within the DIM paradigm, but meaningful constraints require direct confrontation with data. Because the key phenomena, such as NOs and SOs, unfold on day-to-week timescales, continuous monitoring is essential and rarely achievable with a single facility. Leveraging KL Dra’s unusually favourable observability and frequent outburst activity, we assembled a dataset that is difficult to replicate for any other system: nearly a decade of ground-based monitoring from ZTF, ASAS-SN, and ATLAS, combined with three sets of year-long, multi-sector, and near-continuous TESS coverage. We fit empirical functional models to the high-state components of the TESS light curves (SO precursor, rise, plateau, decay, rebrightenings, and NOs). We then anchored and scaled the ground-based light curves to TESS on a common flux baseline, enabling a quantitative characterisation of inter-SC behaviour in the context of KL Dra’s long-term evolution. The main conclusions of this work are:

\begin{itemize}

    \item We characterised 49 SCs within a $\sim10.7$ yr span, including a continuous sequence of 34 SCs across $\sim5.7$ yr. The mean SC duration is $60.4 \pm 0.1$ d, spanning $43.3 \pm 0.9$ to $77.3 \pm 0.8$ d. An O-C analysis of SO start times reveals three distinct regimes (two decreasing trends separated by an increasing trend), indicating that successive SCs are not fully independent and suggesting a longer-term modulation (``hypercycle'') in the system’s outburst clock.

    \item Rebrightenings exhibit a characteristic morphology (a cluster of short, rising events followed by a longer, steadier event) and a mean duration of $9.7 \pm 0.2$ d (range $5.2-18.2$ d). We find a statistically significant anti-correlation (Spearman coefficient of -0.74) between rebrightening duration and SC duration that flattens towards the long-duration end.

    \item Every TESS-observed SO is best fit with a  precursor. The intrinsic SO lasts $5.67 \pm 0.03$ d on average, with typical phase durations of $0.95 \pm 0.02$ d for the precursor, $0.36 \pm 0.01$ d for the rise-to-plateau, $3.68 \pm 0.06$ d for the plateau, and $0.66\pm0.04$ d for the decay. We find tentative correlations between plateau/decay amplitudes and SC duration below 70 d, and tentative evidence for an inverse relation between plateau duration and SC duration (also below 70 d). 

    \item KL Dra displays 3-4 large amplitude NOs per SC with a mean duration of $1.17 \pm 0.01$ d. Within an SC, NO amplitude, rise timescale, and total duration generally increase from the first to subsequent NOs (the precursor continues the trend of increasing amplitude and rise duration). NO asymmetry evolves within the SC: the first NO is the most asymmetric (faster rise than decay), whereas later NOs are more symmetric ($f_r \sim 0.4-0.6$). Mini-NOs occur at various phases of the SC. Although we do not find a uniform trend for the inter-NO quiescent intervals, they appear to track SC length.

    \item Several of our constraints challenge a strictly tidal-thermal instability interpretation and point to additional ingredients such as time-dependent enhancements of the mass-transfer rate. Additionally, other model parameters such as helium-specific opacity effects, potential inner-disc truncation, and irradiation make direct comparison between systems difficult.
    
\end{itemize}

This is the first time the outburst morphology of an AM CVn system has been quantified in such detail. Placing the outburst evolution within the context of long-term SC behaviour has likewise been largely unexplored, even for dwarf novae. Our dataset and characterisation represent an important step towards understanding accretion phenomena in compact binaries and provide a reference benchmark against which disc-instability models can be tested.

\section{Acknowledgements} \label{sec:acknow}

L.S.M. thanks Richard Pomeroy for guidance in accessing ZTF data. 
L.S.M. and L.R.S. acknowledge support from NASA grant NNH21ZDA001N-5135.  
COH is supported by NSERC Discovery Grant RGPIN-2023-04264, and Alberta Innovates Advance Program 242506334. MPM is partially supported by the Swiss National Science Foundation IZSTZ0\_216537 and by UNAM PAPIIT-IG101224. 

This paper includes data collected by the TESS mission, which are publicly available from the Mikulski Archive for Space Telescopes (MAST). Funding for the TESS mission is provided by NASA's Science Mission directorate. 

Based on observations obtained with the Samuel Oschin Telescope 48-inch and the 60-inch Telescope at the Palomar Observatory as part of the Zwicky Transient Facility project. ZTF is supported by the National Science Foundation under Grants No. AST-1440341 and AST-2034437 and a collaboration including current partners Caltech, IPAC, the Oskar Klein Center at Stockholm University, the University of Maryland, University of California, Berkeley , the University of Wisconsin at Milwaukee, University of Warwick, Ruhr University, Cornell University, Northwestern University and Drexel University. Operations are conducted by COO, IPAC, and UW. 

We thank Las Cumbres Observatory and its staﬀ for their continued support of ASAS-SN. ASAS-SN is funded in part by the Gordon and Betty Moore Foundation through grants GBMF5490 and GBMF10501 to the Ohio State University, and also funded in part by the Alfred P. Sloan Foundation grant G-2021-14192. Development of ASAS-SN has been supported by NSF grant AST-0908816, the Mt. Cuba Astronomical Foundation, the Center for Cosmology and AstroParticle Physics at the Ohio State University, the Chinese Academy of Sciences South America Center for Astronomy (CAS-SACA),and the Villum Foundation.

This work has made use of data from the Asteroid Terrestrial-impact Last Alert System (ATLAS) project. The Asteroid Terrestrial-impact Last Alert System (ATLAS) project is primarily funded to search for near earth asteroids through NASA grants NN12AR55G, 80NSSC18K0284, and 80NSSC18K1575; byproducts of the NEO search include images and catalogues from the survey area. This work was partially funded by Kepler/K2 grant J1944/80NSSC19K0112 and HST GO-15889, and STFC grants ST/T000198/1 and ST/S006109/1. The ATLAS science products have been made possible through the contributions of the University of Hawaii Institute for Astronomy, the Queen’s University Belfast, the Space Telescope Science Institute, the South African Astronomical Observatory, and The Millennium Institute of Astrophysics (MAS), Chile.

\appendix
\counterwithin{figure}{section}

\section{Supercycles light curves and modelling} \label{app:SCs}

We present the light curves and fitted models for the SCs that meet our inclusion criteria (Figure~\ref{fig:allSCs}): (i) start and end times are constrained, and (ii) intra-SC coverage is sufficient to rule out additional SOs. As noted in Section~\ref{subsec:Groundlcs}, the ground-based light curves (ZTF, ASAS-SN, ATLAS) were scaled to the TESS amplitudes so that all curves share the same y-axis. The fitting of the SO, rebrightening, and NO components is described in Section~\ref{subsec:outact}.  The sample contains six sets of contiguous SC measurements (see the sequences connected by lines in Figure~\ref{fig:SCdur}).

Our naming scheme is SC-group-index (displayed top-right in Figure~\ref{fig:allSCs}). Groups follow increasing JD: a, b, c, d, e, f, and indices increase within each group. Groups a and b contain 4 SCs each, groups c and d contain 2 SCs each, group e contains 3 SCs, and group f is a contiguous sequence of 34 SCs. The only sparsely sampled SC included is SC-f-11. The gap is consistent with quiescence as adding a missed SO would force unreasonable durations for two SCs. Supporting this, the measured rebrightening duration for SC-f-11 matches the long-duration trend in Figure~\ref{fig:combined-figures}, arguing against a short duration SC.

\begin{figure*}[h]
    \centering
    \begin{subfigure}[b]{0.48\linewidth}
        \centering
        \includegraphics[width=\linewidth]{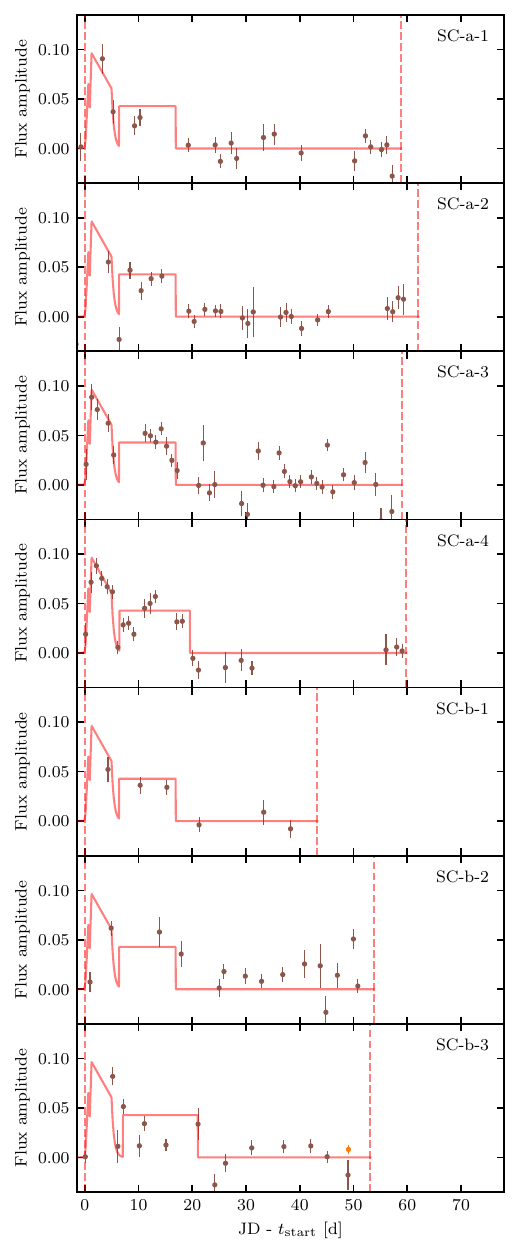}
        \label{fig:subfig1}
    \end{subfigure}
    \hfill
    \begin{subfigure}[b]{0.48\linewidth}
        \centering
        \includegraphics[width=\linewidth]{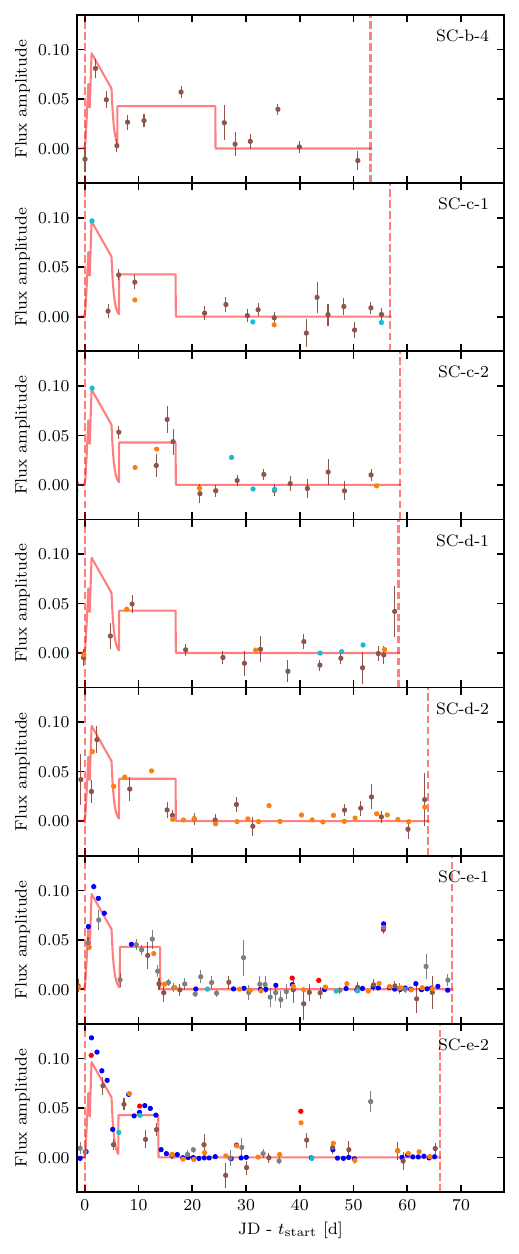}
        \label{fig:subfig2}
    \end{subfigure}
    \caption{Light curves and models for all SCs included in this study, shown as a function of time from the SO onset. The colour/symbol scheme follows Figure~\ref{fig:alldata} and the panel layout follows Figure~\ref{fig:SC_selected}. TESS (2 hour bins) = white circles with black edge; ZTF $g/i$ = blue/red; ASAS-SN $g/V$ = grey/brown; ATLAS $o/c$ = orange/cyan. The red solid curve is the best-fit model; the red dashed line marks the end of the SC. Vertical solid lines indicate NOs and vertical dotted lines indicate mini-NOs. The label of each SC appears in the upper-right corner. The TESS sector is marked by a gray horizontal band, with the sector number indicated at the beginning and end of the corresponding time interval.}
    
    \label{fig:allSCs}
\end{figure*}

\begin{figure*}[h]
    \ContinuedFloat

    \centering
    \begin{subfigure}[b]{0.48\linewidth}
        \centering
        \includegraphics[width=\linewidth]{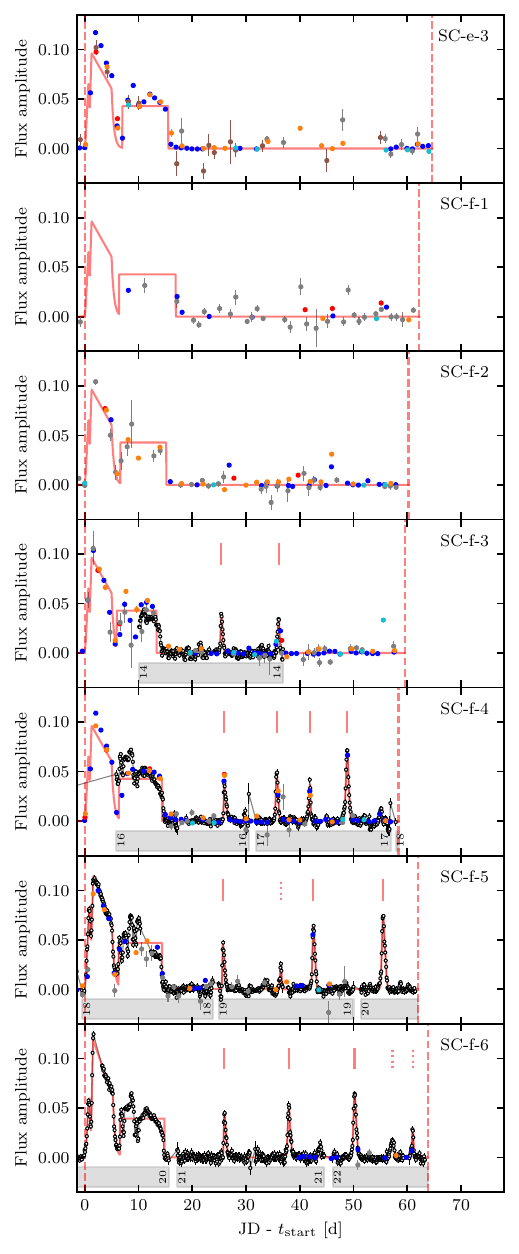}
        \label{fig:subfig3}
    \end{subfigure}
    \hfill
    \begin{subfigure}[b]{0.48\linewidth}
        \centering
        \includegraphics[width=\linewidth]{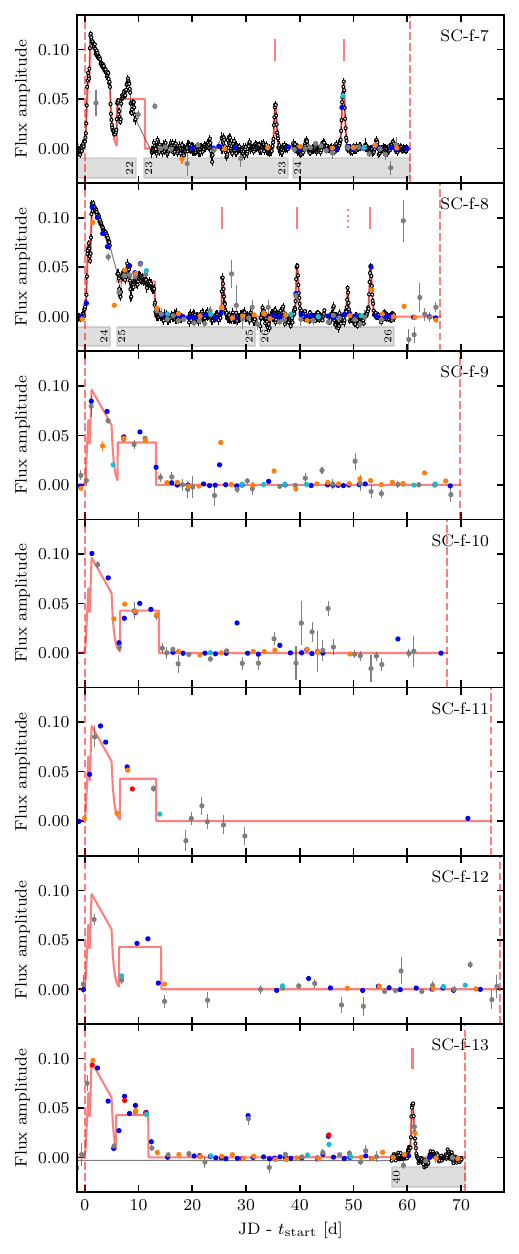}
        \label{fig:subfig4}
    \end{subfigure}
    \caption{Continuation of Figure \ref{fig:allSCs}.}
    \label{fig:main4}
\end{figure*}

\begin{figure*}[h]
    \ContinuedFloat
    \centering
    \begin{subfigure}[b]{0.48\linewidth}
        \centering
        \includegraphics[width=\linewidth]{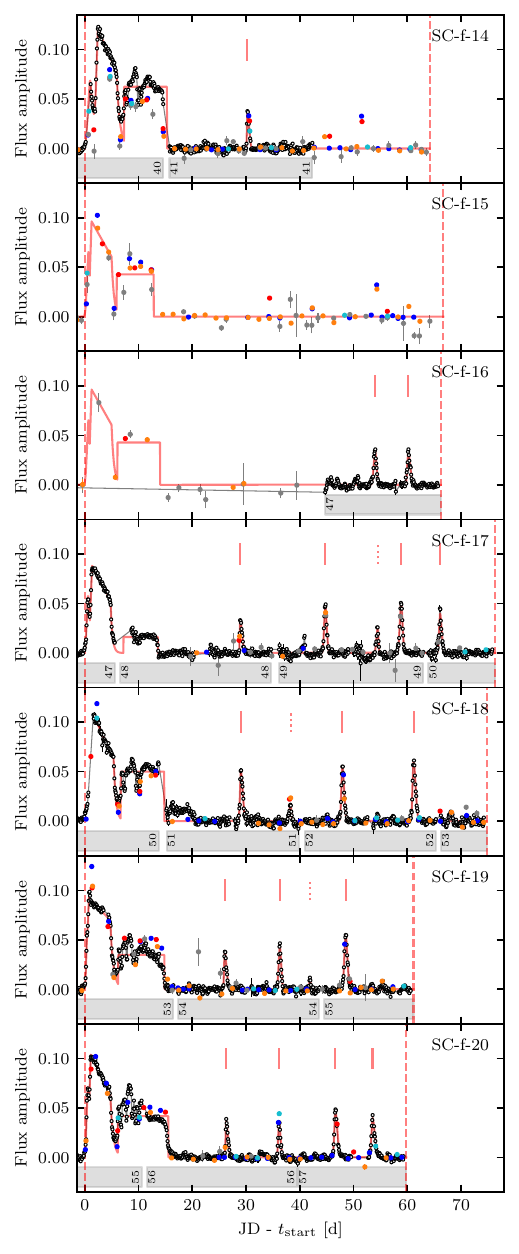}
        \label{fig:subfig5}
    \end{subfigure}
    \hfill
    \begin{subfigure}[b]{0.48\linewidth}
        \centering
        \includegraphics[width=\linewidth]{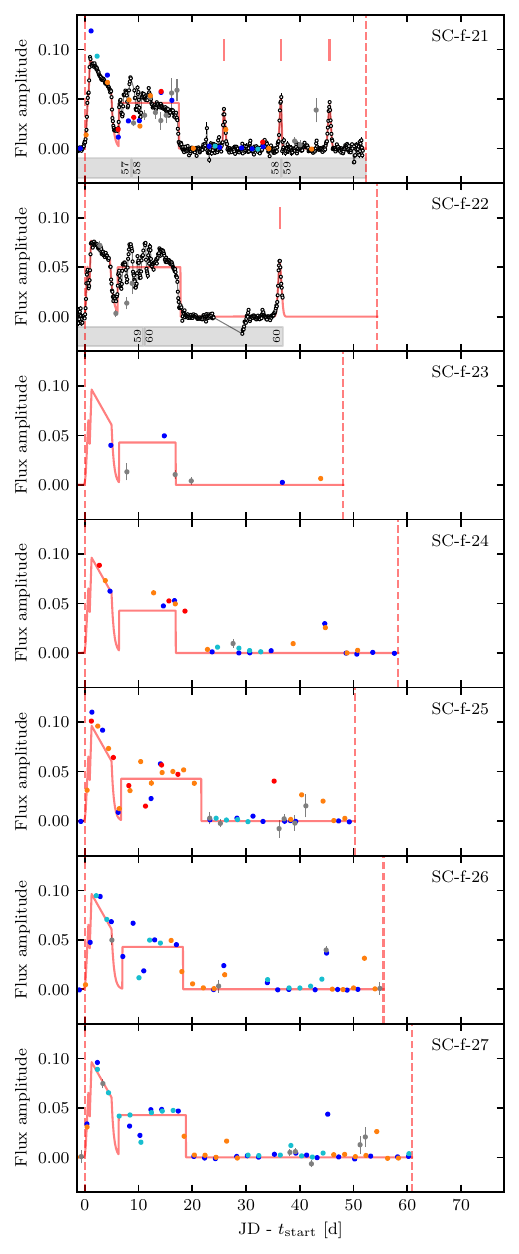}
        \label{fig:subfig6}
    \end{subfigure}
    \caption{Continuation of Figure \ref{fig:allSCs}.}
    \label{fig:main6}
\end{figure*}

\begin{figure}[h]
    \ContinuedFloat
    \centering
    \begin{subfigure}[b]{\linewidth}
        \centering
        \includegraphics[width=\linewidth]{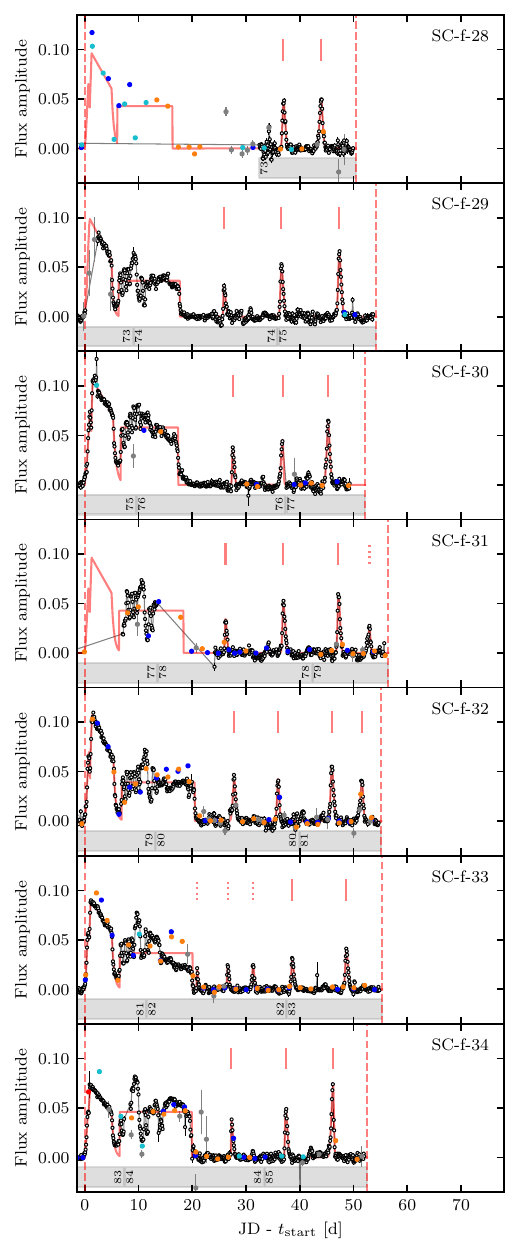}
        \label{fig:subfig5}
    \end{subfigure}
    \caption{Continuation of Figure \ref{fig:allSCs}.}
\end{figure}



\bibliographystyle{paslike}
\bibliography{references.bib}


\end{document}